\documentclass[journal]{IEEEtran}
\usepackage{amsmath,amssymb,amsfonts}
\usepackage{amsthm}
\newtheorem{theorem}{Theorem}
\usepackage{algorithm}
\usepackage{algpseudocode}
\usepackage{array}
\usepackage[caption=false,font=normalsize,labelfont=sf,textfont=sf]{subfig}
\usepackage{xcolor}
\usepackage{textcomp}
\usepackage{stfloats}
\usepackage{url}
\usepackage{verbatim}
\usepackage{graphicx}
\usepackage{cite,comment}
\usepackage{epstopdf}
\usepackage{epsfig}
\usepackage{hyperref}
\usepackage{cuted,color}
\usepackage{epstopdf} 
\usepackage{subfig}
\newtheorem{corollary}{Corollary}

\newcommand{\mv}[1]{\boldsymbol{#1}}

\begin{document}

\title{Movable Antenna Empowered Near-Field Sensing via Antenna Position Optimization}

\author
{Yushen~Wang, Weidong~Mei,~\IEEEmembership{Member,~IEEE,} Xin~Wei, Ya~Fei~Wu,~\IEEEmembership{Member,~IEEE,} Zhi~Chen,~\IEEEmembership{Senior~Member,~IEEE,} and~Boyu~Ning,~\IEEEmembership{Member,~IEEE} 
\thanks{
Part of this paper has been presented at the IEEE International Conference on Communications (ICC) workshop, Montreal, Canada, 2025 \cite{wang2025antenna}.

Yushen Wang is with the Glasgow College, University of Electronic Science and Technology of China, Chengdu 611731, China (e-mail: micro7386@gmail.com).
 
Weidong Mei, Xin Wei and Zhi Chen are with the National Key Laboratory of Wireless Communications, University of Electronic Science and Technology of China, Chengdu 611731, China (e-mail: wmei@uestc.edu.cn, xinwei@std.uestc.edu.cn, chenzhi@uestc.edu.cn).

Ya Fei Wu is with the EHF Key Laboratory of Fundamental Science, School of Electronic Science and Engineering, University of Electronic Science and Technology of China, Chengdu 611731, China (email: wuyafei@uestc.edu.cn).

Boyu Ning is with the Department of Communication Systems, KTH Royal Institute of Technology, Stockholm, Sweden (email: boydning@outlook.com).
}}



\maketitle

\begin{abstract}
Movable antenna (MA) technology exhibits great promise for enhancing the sensing capabilities of future sixth-generation (6G) networks due to its capability to alter antenna array geometry. With the growing prevalence of near-field propagation at ultra-high frequencies, this paper focuses on the application of one-dimensional (1D) and two-dimensional (2D) MA arrays for near-field sensing to jointly estimate the angle and distance information about a target. First, for the 1D MA array scenario, to gain insights into MA-enhanced near-field sensing, we investigate two simplified cases with only angle-of-arrival (AoA) or distance estimation, respectively, assuming that the other information is already known. The worst-case Cramer–Rao bounds (CRBs) on the mean square errors (MSEs) of the AoA estimation and the distance estimation are derived in these two cases. Then, we jointly optimize the positions of the MAs within the 1D array to minimize these CRBs and derive their closed-form solutions, which yield an identical array geometry to MA-enhanced far-field sensing. For the more challenging joint AoA and distance estimation, since the associated worst-case CRB is a highly complex and non-convex function with respect to the MA positions, a discrete sampling-based approach is proposed to sequentially update the MA positions and obtain an efficient suboptimal solution. Furthermore, we investigate the worst-case CRB minimization problems for a 2D MA array under various conditions and extend our proposed algorithms to solve them efficiently. Numerical results demonstrate that the proposed MA-enhanced near-field sensing scheme dramatically outperforms conventional fixed-position antennas (FPAs). Moreover, the joint angle and distance estimation results in a different array geometry from that in the individual estimation of angle/distance or far-field sensing.
\end{abstract}


\section{Introduction}
\IEEEPARstart{I}{n} future sixth-generation (6G) wireless systems, substantial advancements in both communication and sensing capabilities are anticipated \cite{liu2022integrated,liu2022survey}. Extensive research efforts have been devoted to enabling 6G networks that not only support ultra-high transmission rates but also provide accurate detection, estimation, and acquisition of environmental information, driven by emerging applications such as smart healthcare, vehicle-to-everything (V2X), and virtual reality (VR) \cite{chowdhury2020wireless,jiang2021road}. In particular, wireless sensing is envisioned to become a key service in future 6G networks.

Wireless sensing systems typically employ antenna arrays at the transmitter to actively emit probing signals over multiple temporal snapshots and receive their echoes for estimating key target parameters such as time-of-arrival (ToA), Doppler shift, and angle-of-arrival (AoA) \cite{xie2023collaborative,ni2023uplink}. To enhance parameter resolution and beamforming gain, large antenna arrays with extended apertures are commonly deployed at the base station (BS). Meanwhile, future wireless systems are expected to operate at higher frequency bands (e.g., Terahertz) to access broader bandwidths. The combination of larger apertures and higher frequencies necessitates the adoption of a near-field spherical-wave propagation model for both communication and sensing applications\cite{ning2023beamforming,wang2023near,liu2023near}.
Unlike conventional far-field sensing that relies solely on angular information, near-field sensing inherently couples both angular and spatial parameters in the received signals, which can be jointly exploited for high-resolution target detection, localization, and tracking. Consequently, near-field wireless sensing has recently garnered growing interest and demonstrated significant potential\cite{wang2018unified,friedlander2019localization,dardari2022near}. However, most existing works on near-field sensing employ fixed-position antennas (FPAs), whose fixed geometry may limit sensing performance. For example, a compact half-wavelength FPA array usually suffers from a limited aperture, resulting in low sensing resolution. While a sparse FPA array can improve sensing resolution by enlarging the aperture, this often introduces considerable grating lobes that cause parameter‑estimation ambiguities. By contrast, MAs can reshape the array geometry according to the sensing task or target region, which provides additional spatial degrees of freedom (DoFs) to both enlarge the aperture and suppress ambiguous parametric correlations.

To tackle this issue, this paper proposes the use of movable antenna (MA) technology for near-field wireless sensing. In contrast to FPAs, MAs can dynamically adjust their positions within a given region via mechanical actuators or electronic control for various purposes\cite{wong2021fluid,zhu2025tutorial,ning2025movable,zhu2024movable,zhu2024modeling}. First, antenna positions can be adaptively optimized based on instantaneous channels to avoid/achieve deep fading for enhancing/suppressing desired/undesired signals. In \cite{zhu2024antenna,xiao2024multiuser,feng2024weighted,ma2024mimo,zhang2025movable}, MAs were shown to be able to effectively improve the rate performance of single-/multi-user multi-antenna systems by achieving more favorable channel conditions and higher spatial multiplexing gains. Specifically, the authors in \cite{zhu2024antenna} and \cite{xiao2024multiuser} investigated antenna position optimization problems in user- and BS-side MA-enhanced multi-user communication systems, respectively. The authors in \cite{feng2024weighted} explored a more general scenario in which both the BS and the users are equipped with MAs and formulated a weighted sum-rate maximization problem. In addition, the authors in \cite{ma2024mimo} extended the above results to a multiple-input multiple-output (MIMO) setup. It was shown in \cite{ma2024mimo} that antenna position optimization can improve the MIMO channel power while decreasing the channel condition number, thereby significantly boosting the MIMO channel capacity compared with FPAs. Furthermore, the authors in \cite{zhang2025movable} investigated an array-level MA architecture combined with hybrid beamforming for multi-user communications. Other recent contributions in the literature have further explored the application of MAs for physical-layer security \cite{mei2024posistion,hu2024secure}, over-the-air computation \cite{li2025over,zhang2024fluid}, cognitive radio \cite{wei2024joint}, relaying systems \cite{li2025movable}, non-orthogonal multiple access (NOMA) \cite{li2024sum}, intelligent reflecting surface (IRS)-aided wireless communications \cite{wei2025movable}, etc.

Second, by adjusting antenna positions, the array geometry can be reshaped to alter the spatial correlation among steering vectors corresponding to different angles, thereby achieving more flexible array signal processing. For example, in \cite{zhu2023movable}, the authors demonstrated that MA-enabled arrays can simultaneously achieve beam nulling towards undesired directions and full array gain towards desired directions by dynamically repositioning antennas. Moreover, the authors in \cite{ma2024multi,liu2025movable} showed that MAs can also facilitate multi-beam forming by maximizing the spatial correlations of array responses corresponding to different angles. Furthermore, the authors in \cite{wang2025movable} studied MA-enhanced wide-beam coverage within a given spatial region and proposed an efficient frequency modulation continuous wave (FMCW)-based design to solve the corresponding antenna position optimization problem.

Third, in terms of wireless sensing, MAs can be exploited to enlarge the apertures of antenna arrays compared with FPA arrays. This creates more favorable propagation conditions for target localization and also enhances the angle and distance estimation resolution. In particular, the authors in \cite{ma2024movable} derived the Cramer-Rao bound (CRB) on the angle estimation in an MA-enhanced sensing system and optimized the MA positions to minimize this CRB. The authors in \cite{shao2025exploiting} later extended the results in \cite{ma2024movable} to more general six-dimensional MA (6DMA)-aided sensing systems. Furthermore, in \cite{ma2025movable, lyu2025movable}, the authors investigated the MA position optimization problems for integrated sensing and communications (ISAC) and showed that MAs can greatly improve the sensing-communication performance trade-off. However, all of the above works consider MA-enhanced far-field sensing with angle estimation only, leaving the challenge of joint angle and distance estimation in near-field scenarios unresolved. To the best of our knowledge, there is no existing work focusing on MA-enhanced near-field sensing so far.

To fill in this gap, this paper investigates an MA-enhanced near-field wireless sensing system, aiming to estimate the angle and distance information of a target in the near-field region of a one-dimensional (1D) linear array and a two-dimensional (2D) planar array, respectively. The main contributions of this paper are summarized as follows: 
\begin{enumerate}
    \item For the 1D linear MA array, to gain insights, we first investigate the individual estimation of the angle-of-arrival (AoA) and the distance of the target via the multiple signal classification (MUSIC) algorithm, assuming that the other parameter is already known. The CRBs on the mean square errors (MSEs) of the AoA and distance estimations are derived, respectively. Since the CRBs are jointly determined by the corresponding estimators, we aim to minimize the worst-case CRBs with respect to (w.r.t.) the estimators by optimizing the antenna position vector (APV). Closed-form solutions to the CRB minimization problems are derived for both cases; notably, they yield an identical array geometry to that in the far-field sensing scenario. Furthermore, we proceed to the general case of joint AoA and distance estimation via the two-dimensional (2D) MUSIC algorithm, and derive the corresponding worst-case sum of the CRBs for both estimators. To tackle this more challenging optimization problem, a discrete sampling-based algorithm is proposed, where the movement region is discretized into a set of sampling points and the positions of the MAs are sequentially updated until convergence.
    \item Next, we extend the above analytical framework to the more challenging case of a 2D planar MA array. In this context, we derive and minimize the worst-case (sum) CRBs for three scenarios: individual 2D elevation-azimuth estimation, individual distance estimation, and joint three-dimensional (3D) AoA and distance estimation. Due to the high non-convexity of these worst-case CRBs in the 2D MA case, the discrete sampling-based algorithm is adopted to alleviate the structural complexity and obtain suboptimal antenna position matrices (APMs). Extensive simulation results demonstrate the superiority of the proposed scheme over conventional FPA-based benchmarks with both half-wavelength and sparse antenna spacings. 
\end{enumerate}

The rest of this paper is organized as follows. Section \ref{sec_sysmodel} presents the system model for 1D MA arrays. Section \ref{sec_1D} derives the worst-case CRBs and presents the proposed algorithms for CRB minimization in individual and joint parameter estimation with 1D MA arrays. Section \ref{sec_2D} extends the results in Section \ref{sec_1D} to 2D MA arrays. Numerical results and discussions are provided in Section \ref{sec_numerical}. Finally, Section \ref{sec_conclusion} concludes this paper.

\textit{Notations}: Boldface lower- and upper-case letters represent vectors and matrices, respectively. The conjugate, transpose, conjugate transpose, and trace are represented by $(\cdot)^*$, $(\cdot)^\top$, $(\cdot)^\mathsf{H}$, and $\text{Tr}(\cdot)$, respectively. The sets of $(N_1 \times N_2)$-dimensional real and complex matrices are denoted by $\mathbb{R}^{N_1 \times N_2}$ and $\mathbb{C}^{N_1 \times N_2}$, respectively. $\|\mv{v}\|$ denotes the $l_2$-norm of a vector $\mv{v}$. $|{\cal A}|$ denotes the cardinality of a set ${\cal A}$ and ${\cal A}\backslash {\cal B}$ denotes the subtraction of set ${\cal B}$ from set ${\cal A}$. The expectation operator is denoted by $\mathbb{E}\{\cdot\}$. The Hadamard product is denoted by $\odot$. $\mv{I}_N$ denotes the $N$-dimensional identity matrix and $\mathbf{1}_N$ denotes the $N$-dimensional all-one vector.

\begingroup
\allowdisplaybreaks
\section{System Model}\label{sec_sysmodel}
\begin{figure}[t]
    \centering
    \includegraphics[scale=0.35]{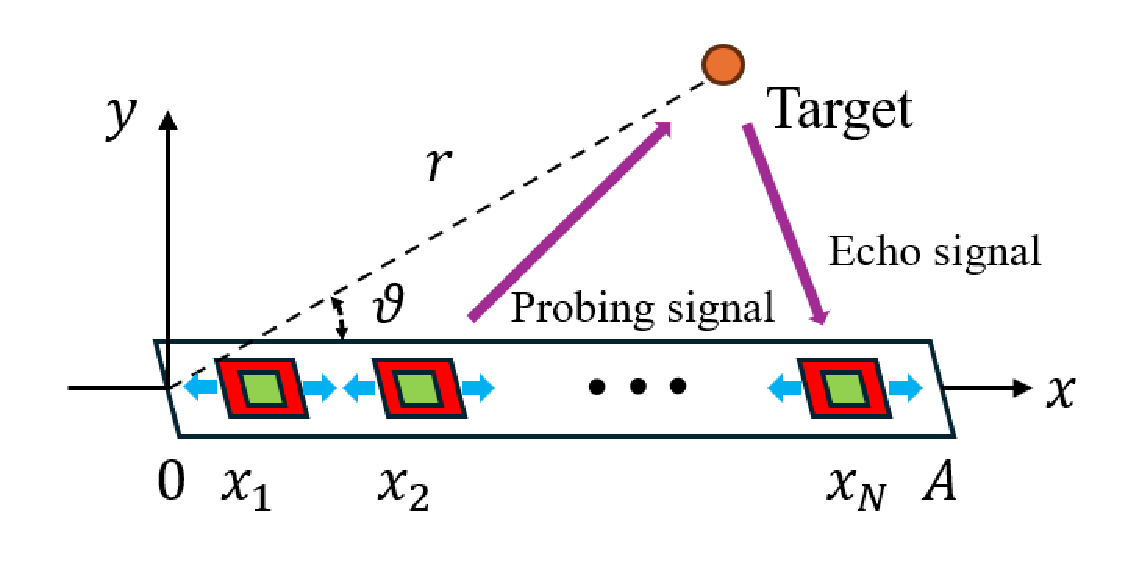}
    \vspace{-9pt}
    \caption{1D MA array for near-field target localization.}
    \label{1D_sysmodel}
    \vspace{-9pt}
\end{figure}
As shown in Fig. \ref{1D_sysmodel}, we first consider a 1D near-field wireless sensing system with \( N \) MAs to estimate the angular/spatial parameter(s) of a target. The positions of the MAs can be flexibly adjusted within a linear segment of length \( A \). Denote the position of the \( n \)-th MA (\( n \in {\cal N} \triangleq \{1, 2, \ldots, N\} \)) by \( x_n \in [0, A] \), and the APV of all \( N \) MAs by $\mv{x} \triangleq [x_1, x_2, \ldots, x_N]^\top \in \mathbb{R}^N$. Without loss of generality, we assume that \( 0 \leq x_1 < x_2 < \cdots < x_N \leq A \). Therefore, the effective aperture of the MA array can be represented as $D = x_N - x_1$. We assume that the target is located in the near-field region of the linear array but outside its reactive region, which means that the distance between the target and any position within the MA array is between the Fresnel distance and the Rayleigh distance, which are respectively given by $R_{FS} \triangleq (\frac{A^4}{8\lambda})^{\frac{1}{3}}$ \cite{selvan2017fraunhofer} and $R_{RL} \triangleq \frac{2A^2}{\lambda}$ \cite{liu2023near}, where $\lambda$ is the signal wavelength. 

The MA positions are pre‑optimized based on the possible target region and remain fixed during the entire sensing process. At the beginning of the sensing process, the MA array transmits sensing signals and receives the echoes reflected from the target, which is assumed to remain static throughout the process\cite{wang2023near, liu2022cramer}. To characterize the near-field channel from the antenna array to the target, we adopt the uniform spherical wave (USW) channel model in \cite{liu2023near}, where the channel coefficients have identical amplitudes across all MAs, while their phases vary across them. As depicted in Fig. \ref{1D_sysmodel}, we denote the physical steering angle between the origin and the target as $\theta$, with $\theta \in [\theta_{\text{min}}, \frac{\pi}{2}]$, where $\theta_{\text{min}} > 0$ is the prescribed lower bound of the steering angle.\footnote{Note that we only consider $\theta \le \frac{\pi}{2}$ here. This is because in the case of $\theta >\frac{\pi}{2}$, we can redefine the origin as the ending position of the MA array, which yields the same sensing result thanks to symmetry.} Accordingly, the directional cosine of the AoA is defined as $u = \cos\theta \in [0, u_{\text{max}}]$, where $u_{\text{max}}=\cos\theta_{\text{min}}$ is its upper bound. Denote $\mv{s}_n = [x_n,0]^\top$ as the coordinate of the \( n \)-th MA, and $r \in [r_{\text{min}}, r_{\text{max}}]$ as the distance between the origin and the target, where $r_{\text{min}}$ and $r_{\text{max}}$ are the prescribed lower- and upper-bounds of the distance, respectively. Thus, the coordinate of the target is given by $\mv{r} = [r \cos \theta, r \sin \theta]^\top$. Then, the distance from the \( n \)-th MA to the target can be expressed as a function of the APV $\mv{x}$ and the target parameter vector denoted by $\mv{\eta} = [u, r]^\top$, i.e.,
\begin{align}
\label{Taylor}
r_n(x_n, \mv{\eta}) = \|\mv{r}-\mv{s}_n\| &= \sqrt{r^2-2\mv{r}^\top\cdot \mv{s}_n+\|\mv{s}_n\|^2} \\ 
&= \sqrt{r^2-2 x_n u r +x_n^2}. \notag
\end{align}
By invoking the Fresnel approximation for the near-field model\cite{liu2023near}, the distance in \eqref{Taylor} can be approximated as the second-order Taylor expansion based on $\sqrt{1+x} \approx 1+\frac{1}{2}x-\frac{1}{8}x^2$ with $x = (-2\mv{r}^\top \cdot \mv{s}_n+\|\mv{s}_n\|^2)/r^2$, i.e.,
\begin{align}
r_n(x_n, \mv{\eta}) &\approx r-x_n u+\frac{x_n^2(1-u^2)}{2r}.
\end{align}
Let ${\beta}_0$ denote the free-space path loss between the MA array and the target. Then, the channel coefficient between the \( n \)-th MA and the target is given by
\begin{align}
h_n(x_n, \mv{\eta}) &= \sqrt{{\beta}_0} \exp\left(-j\frac{2\pi}{\lambda}r_n(x_n, \mv{\eta})\right) \notag \\ &= \beta \exp\left(j\frac{2\pi}{\lambda}\big(x_n u-\frac{x_n^2(1-u^2)}{2r}\big)\right),
\label{1D_channel}
\end{align}
where $\beta = \sqrt{{\beta}_0} \exp(-j\frac{2\pi}{\lambda}r)$ is the complex channel gain. As a result, the echoed LoS channel vector can be written as
\begin{align}
\mv{h}(\mv{x}, \mv{\eta}) &= [h_1(x_1, \mv{\eta}), h_2(x_2, \mv{\eta}), \dots, h_N(x_N, \mv{\eta})]^\top \\ &= \beta \mv{\alpha}(\mv{x}, \mv{\eta}) \in \mathbb{C}^N, \notag
\end{align}
where $\mv{\alpha}(\mv{x}, \mv{\eta})$ denotes the near-field steering vector of the MA array. In this paper, we aim to estimate the target parameters by properly setting the APV $\mv{x}$, as detailed next.

To characterize the estimation accuracy of the antenna array, we adopt the CRBs on the estimators, which also serve as theoretical lower bounds on their estimation MSEs. Hence, we aim to optimize the MA positions to minimize the CRBs on the estimators. Note that compared with far-field sensing involving only the angular domain, near-field sensing involves both angular and spatial information, thus facilitating target localization \cite{ma2025movable, lyu2025movable}. In the following, to gain insights into the effects of the antenna positions on the target sensing accuracy in the near-field, we consider the following three cases in the next section.
\begin{enumerate}
    \item \textbf{Estimation of AoA only for the 1D MA array (Case 1.1)}: $r$ is known while $u$ is unknown;
    \item \textbf{Estimation of distance only for the 1D MA array (Case 1.2)}: $u$ is known while $r$ is unknown;
    \item \textbf{Joint estimation of AoA and distance for the 1D MA array (Case 1.3)}: both $u$ and $r$ are unknown.
\end{enumerate}

\section{MA-Enhanced Near-Field Sensing For 1D Arrays}
\label{sec_1D}
\subsection{AoA Estimation in Case 1.1}
In Case 1.1, we assume that the distance from the target to the origin of the MA array is already known and denoted as $r^{\star}$, such that only the AoA $u$ needs to be estimated.

For any given APV \(\mv{x}\), the received signals within a number of consecutive snapshots can be collectively adopted to estimate the AoA of the target via the multiple signal classification (MUSIC) algorithm. Let $T$ denote the total number of snapshots. The received echo signal at the MA array in the \(t\)-th snapshot (\(t = 1, 2, \ldots, T\)) is expressed as
\begin{equation}
    \mv{y}_t = \mv{h}(\mv{x}, u)s_t + \mv{w}_t,
\end{equation}
where \(s_t\) represents the sensing signal with \(\mathbb{E}\{\lvert s_t\rvert^2\} = P\), with $P$ denoting the transmit power, and \(\mv{w}_t \sim \mathcal{CN}(\mv{0}, \sigma^2\mv{I}_N)\) is the receiver noise following the circularly symmetric complex Gaussian (CSCG) distribution, with $\sigma^2$ denoting the average noise power.

To estimate the AoA, the received signals across the \(T\) snapshots are arranged into the following matrix as
\begin{equation}
\mv{Y} \triangleq
    \begin{bmatrix}
    \mv{y}_1, \mv{y}_2, \ldots, \mv{y}_T
    \end{bmatrix}
    = \mv{h}(\mv{x}, u)\mv{s}^\top + \mv{W},
\end{equation}
where \(\mv{s} \triangleq [s_1, s_2, \ldots, s_T]^\top \in \mathbb{C}^T\) and \(\mv{W} \triangleq [\mv{w}_1, \mv{w}_2, \ldots, \mv{w}_T] \in \mathbb{C}^{N \times T}\). Therefore, the covariance matrix of \(\mv{Y}\) can be given by
\begin{equation}
    \mv{R_Y} = \frac{1}{T}\mv{Y}\mv{Y}^\mathsf{H} = \frac{1}{T} \mv{h}(\mv{x}, u)\mv{s}^\mathsf{H}\mv{s}\mv{h}(\mv{x}, u)^\mathsf{H} + \sigma^2\mv{I}_N.
\end{equation}
Based on the procedures of the MUSIC algorithm, we can perform the singular value decomposition (SVD) of \(\mv{R_Y}\) as
\begin{equation}
    \mv{R_Y} = 
    \begin{bmatrix}
    \mv{u}_{\mv{s}}, \mv{U}_{\mv{w}}
    \end{bmatrix}
    \begin{bmatrix}
    \gamma_{\mv{s}} & \\
    \ & \mv\Gamma_{\mv{w}}
    \end{bmatrix}
    \begin{bmatrix}
    \mv{u}_{\mv{s}}^\mathsf{H} \\
    \mv{U}_{\mv{w}}^\mathsf{H}
    \end{bmatrix},
\end{equation}
where \(\mv{u}_{\mv{s}} \in \mathbb{C}^N\) and \(\mv{U}_{\mv{w}} \in \mathbb{C}^{N \times (N-1)}\) are the singular vector and matrix of the signal and noise subspaces, respectively, \(\gamma_{\mv{s}}\) denotes the singular value of the signal subspace, and \(\mv\Gamma_{\mv{w}} \in \mathbb{R}^{(N-1)\times(N-1)} \) represents a diagonal matrix with the singular values of the noise subspace on the diagonal. Since \( \mv{\alpha}(\mv{x}, u) \) is orthogonal to \(\mv{U}_{\mv{w}} \), while \( \mv{\alpha}(\mv{x}, \tilde{u}) \) is non-orthogonal to \( \mv{U}_{\mv{w}} \) for $\tilde u \ne u$, we have $\mv{\alpha}(\mv{x}, u)^\mathsf{H}\mv{U}_{\mv{w}}\mv{U}_{\mv{w}}^\mathsf{H}\mv{\alpha}(\mv{x}, u) = 0 \quad \text{and} \quad \mv{\alpha}(\mv{x}, \tilde{u})^\mathsf{H}\mv{U}_{\mv{w}}\mv{U}_{\mv{w}}^\mathsf{H}\mv{\alpha}(\mv{x}, \tilde{u}) \neq 0$. Hence, there is a peak for the spectrum function $p(\Bar{u}) \triangleq \frac{1}{\mv{\alpha}(\mv{x}, \Bar{u})^\mathsf{H}\mv{U}_{\mv{w}}\mv{U}_{\mv{w}}^\mathsf{H}\mv{\alpha}(\mv{x}, \Bar{u})}$ at \( \Bar{u} = u \), and the estimation of \( u \) is given by
\begin{equation}
    \hat{u} = \arg\max_{\Bar{u} \in [0, u_{\text{max}}]} \frac{1}{\mv{\alpha}(\mv{x}, \Bar{u})^\mathsf{H} \mv{U}_{\mv{w}} \mv{U}_{\mv{w}}^\mathsf{H} \mv{\alpha}(\mv{x}, \Bar{u})},
\end{equation}
which can be solved by performing a 1D search. Then, the AoA estimation MSE can be expressed as 
\begin{equation}
\text{MSE}(u) \triangleq \mathbb{E}\{\lvert u - \hat{u}\rvert^2\},
\end{equation}
and its CRB is given by \cite{kay1993fundamentals,stoica1989music}
\begin{equation}
    \text{CRB}_u(\mv{x}, u) = \frac{\kappa}{F_u(\mv{x}, u)} \le \text{MSE}(u),
\label{1_1_CRBu}
\end{equation}
where
\begin{equation}
\kappa \triangleq \frac{\sigma^2 \lambda^2}{8\pi^2 T P N \lvert\beta\rvert^2},
\label{kappa}
\end{equation}
and
\begin{equation}
    F_u(\mv{x}, u) \triangleq \text{var}(\mv{x})+\frac{2u}{r^\star}\text{cov}(\mv{x},\mv{\tilde{x}})+\frac{u^2}{{r^\star}^2}\text{var}(\mv{\tilde{x}}),
\label{1_1_Fxu}
\end{equation}
where $\tilde{\mv{x}} \triangleq [\tilde{x}_1, \tilde{x}_2, \dots, \tilde{x}_N]^\top \in \mathbb{R}^N$ and $\tilde{x}_n \triangleq x_n^2, n \in {\cal N}$. The variance functions are defined as $\text{var}(\mv{x}) \triangleq \frac{1}{N} \sum_{n=1}^N x_n^2 - \mu(\mv{x})^2$ with $\mu(\mv{x}) = \frac{1}{N} \sum_{n=1}^N x_n$ being the mean of \(\mv{x}\), and $\text{var}(\tilde{\mv{x}}) \triangleq \frac{1}{N} \sum_{n=1}^N \tilde{x}_n^2 - \mu(\tilde{\mv{x}})^2$ with $\mu(\tilde{\mv{x}}) = \frac{1}{N} \sum_{n=1}^N \tilde{x}_n$ being the mean of \(\tilde{\mv{x}}\). The covariance function is defined as $\text{cov}(\mv{x},\tilde{\mv{x}}) \triangleq \frac{1}{N} \sum_{n=1}^N x_n \tilde{x}_n - \mu(\mv{x})\mu(\tilde{\mv{x}})$. The detailed derivations of the CRB in \eqref{1_1_CRBu} are provided in Appendix \ref{appen_1_1}.
\begin{figure}[t]
    \centering
    \includegraphics[scale=0.4]{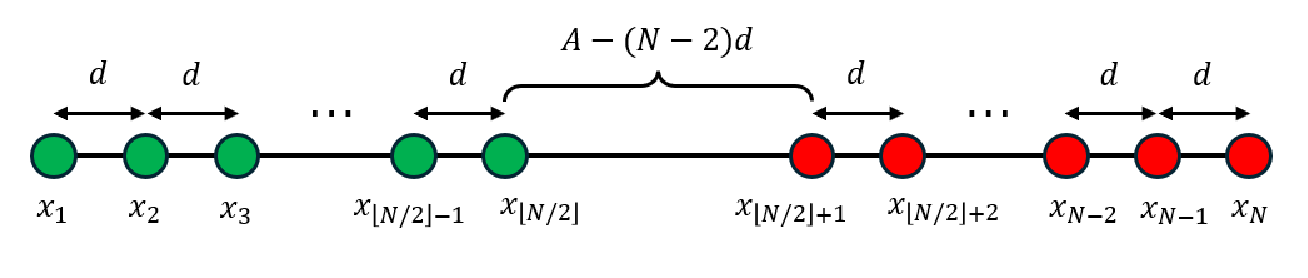}
    \caption{\small Optimal positions of MAs for the 1D MA array in Case 1.1 and Case 1.2.}
    \label{1D_MAarray_Case1_2}
    \vspace{-9pt}
\end{figure}

Our objective is to minimize $\text{CRB}_u(\mv{x}, u)$ by optimizing the APV $\mv{x}$. However, the CRB in \eqref{1_1_CRBu} is dependent on both the APV \(\mv{x}\) and the AoA itself, which are unknown before sensing is performed. To tackle this issue, we aim to minimize the worst-case $\text{CRB}_u(\mv{x}, u)$ for all possible values of the AoA, i.e., $\max_{u} \text{CRB}_u(\mv{x}, u)$. The associated min-max problem can be easily shown to be equivalent to the following max-min problem based on \eqref{1_1_CRBu}, i.e.,
\begin{equation}
    \min_{\mv{x}} \max_{u \in [0, u_{\text{max}}]} \text{CRB}_u(\mv{x}, u) \iff \max_{\mv{x}} \min_{u \in [0, u_{\text{max}}]} F_u(\mv{x}, u).
\label{minmax}
\end{equation}
The associated optimization problem for the right-hand side of \eqref{minmax} can be formulated as
\begin{subequations}
\label{P1-1-star}
\begin{align}
\text{(P1)} \quad
& \max_{\mv{x}} \; F_u(\mv{x}) \triangleq \text{var}(\mv{x})+\frac{2u_{\text{opt}}}{r^\star}\text{cov}(\mv{x},\mv{\tilde{x}})+\frac{u^2_{\text{opt}}}{{r^\star}^2}\text{var}(\mv{\tilde{x}}) \label{P1-1a-star} \\
& \text{s.t.} \quad 0 \le x_n \leq A, \quad n \in {\cal N}, \label{P1-1b-star} \\
& \phantom{\text{s.t.} \quad} \lvert x_i - x_j \rvert \geq d, \quad i \ne j, \quad i,j \in {\cal N}, \label{P1-1c-star}
\end{align}
\end{subequations}
where $d$ denotes the minimum inter-MA distance to avoid mutual coupling, and $u_{\text{opt}}$ is the AoA value that yields the worst-case CRB on the AoA, i.e., $u_{\text{opt}}= \arg\max_{u\in[0, u_{\text{max}}]} \text{CRB}_u(\mv{x}, u)$. By noting that $u \in [0, u_{\text{max}}]$ and that the variance terms $\text{var}(\mv{x})$, $\text{var}(\tilde{\mv{x}})$, and the covariance term $\text{cov}(\mv{x},\tilde{\mv{x}})$ in \eqref{P1-1a-star} are all nonnegative, it can be readily shown that $u_{\text{opt}}=0$. This indicates that the worst-case performance for AoA estimation occurs when the target is in the \textit{broadside} direction of the MA array. As a result, (P1) can be simplified as
\begin{equation}\label{P1-1}
\text{(P1-1)} \quad
\max_{\mv{x}} \; \text{var}(\mv{x}) \quad\text{s.t.} \; \eqref{P1-1b-star}, \eqref{P1-1c-star}. 
\end{equation}
It is worth noting that the optimal MA positions for wide‑beam coverage, as studied in \cite{wang2025movable}, are generally suboptimal for sensing due to the distinct design objectives. In particular, sensing favors a narrow main lobe and low grating lobes to ensure high resolution and unambiguous estimation, rather than generating a wide beam to cover the entire target region.

To maximize the objective function of (P1-1), the MAs should be positioned as dispersively apart as possible, which helps increase the variance term $\text{var}(\mv{x})$, as seen from the theorem below.
\begin{theorem} \label{Th1}
The optimal solution to (P1-1) is given by
\begin{equation}\label{sol_Th1}
x_n^\star = 
\begin{cases} 
(n-1)d, & n = 1, 2, \ldots, \lfloor N/2 \rfloor; \\
A - (N-n)d, & n = \lfloor N/2 \rfloor + 1, \ldots, N.
\end{cases}
\end{equation}
\end{theorem}
\begin{IEEEproof}
Note that problem (P1-1) has the same form as (P1) in \cite{ma2024movable}. Therefore, its optimal APV can be derived by following the same procedures as those in \cite[Appendix A]{ma2024movable}. For brevity, the details are omitted.
\end{IEEEproof}
Moreover, based on Theorem \ref{Th1}, we can also obtain the following corollary.
\begin{corollary} \label{Cor1}
The associated worst-case CRB on the AoA estimation for the optimal APV, i.e., $\text{CRB}_u(\mv{x}^\star,0)$, decreases with $A$ in the order of ${\cal O}(A^{-2})$.
\end{corollary}
\begin{IEEEproof}
Corollary \ref{Cor1} can be readily verified by substituting \eqref{sol_Th1} into the worst-case CRB, i.e., $\text{CRB}_u(\mv{x},0)$. It follows that $\text{CRB}_u(\mv{x}^\star, 0)$ (given in \cite[(52)]{ma2024movable} and omitted here for brevity) decreases with $A$ for $A \geq (N-1)d$ in the order of ${\cal O}(A^{-2})$. This completes the proof.
\end{IEEEproof} 
Theorem \ref{Th1} demonstrates that, to minimize the CRB of the AoA estimation MSE in the near-field, the optimal MA positions are the same as those for the AoA estimation in the far-field, as derived in \cite{ma2024movable}. In particular, the MAs should be divided into two groups, as depicted in Fig. \ref{1D_MAarray_Case1_2}. The first group of MAs is placed at the leftmost end of the 1D MA array, while the other group is at the rightmost end. The performance equivalence between Case 1.1 and far-field scenarios stems from prior knowledge of the target distance, which removes the angle-distance coupling and reduces the AoA-related CRB minimization to maximizing the spatial variance of the antenna positions. Additionally, it can be shown from Corollary \ref{Cor1} that the CRB on the AoA estimation for the optimal APV can be effectively decreased by increasing the length of the MA array, as this results in a larger array aperture, enabling the synthesis of sensing beams with higher angular resolution in the near-field region for a given distance.
\vspace{-9pt}

\subsection{Distance Estimation in Case 1.2}
In this subsection, we consider Case 1.2 where the AoA is already known and denoted as $u^{\star}$. To estimate the distance $r$, we also apply the MUSIC algorithm by leveraging the distance-related information in the signal phase \cite{wang2023near}. For simplicity, the detailed process for distance estimation is omitted here. The associated MSE and CRB are given by
\begin{equation}
    \text{MSE}(r) \geq \text{CRB}_r(\mv{x}, r) = \frac{\kappa}{F_r(\mv{x}, r)},
\label{1_2_CRBr}
\end{equation}
where
\begin{equation}
    F_r(\mv{x}, r) \triangleq \Big(\frac{1-{u^\star}^2}{2r^2}\Big)^2 \text{var}(\tilde{\mv{x}}).
\label{1_2_Fxr}
\end{equation}
The detailed derivations are provided in Appendix \ref{appen_1_2}. Note that the CRB in \eqref{1_2_CRBr} depends on the exact distance $r$. To eliminate its effects, similar to the AoA estimation, we aim to minimize the worst-case CRB among all possible values of distance, which is identical to maximizing the minimum $F_r(\mv{x}, r)$ over $r$. Hence, the corresponding min-max problem can be reformulated as a max-min problem, i.e.,
\begin{equation}
    \min_{\mv{x}} \max_{r \in [r_{\text{min}}, r_{\text{max}}]} \text{CRB}_r(\mv{x}, r) \iff \max_{\mv{x}} \min_{r \in [r_{\text{min}}, r_{\text{max}}]} F_r(\mv{x}, r).
\end{equation}
Denote $r_{\text{opt}}$ as the distance that yields the minimum $F_r(\mv{x}, r)$. Because $F_r(\mv{x}, r)$ is constantly positive due to the variance term $\text{var}(\tilde{\mv{x}})$ and decreases monotonically with $r$, its minimum must occur at the maximum value of $r$, thus leading to $r_{\text{opt}}=r_{\text{max}}$. Then, the optimization problem can be formulated as
\begin{subequations}
\label{P1-2}
\begin{align}
\text{(P1-2)} \quad
& \max_{\mv{x}} \; F_r(\mv{x},r_{\max}) = \Big(\frac{1-{u^\star}^2}{2r^2_{\text{max}}}\Big)^2 \text{var}(\tilde{\mv{x}}) \\
& \text{s.t.} \quad \eqref{P1-1b-star}, \eqref{P1-1c-star}. \notag
\end{align}
\end{subequations}

\begin{algorithm}[t]
\caption{Proposed Algorithm for Solving (P1-3)}
\label{algo_case_1_3}
\begin{algorithmic}[1]
\State \textbf{Input:} $n=1$, ${\cal P}$, and ${\cal P}_{\text{init}}$.
\While{$n \leq N$}
    \State Obtain $x^\star_n$ based on \eqref{arg1} and update $x^{\text{init}}_n \gets x^\star_n$.
    \State Determine ${\cal P}_{n+1}$ based on \eqref{arg1}.
    \State Update $n \gets n+1$.
\EndWhile
\State \textbf{Output:} the optimized APV of all $N$ MAs, i.e., $\mv{x}^\star$.
\end{algorithmic}
\end{algorithm}

\begin{theorem} \label{Th2}
For a sufficiently large array aperture $A$, the optimal solution to (P1-2) is given by \eqref{sol_Th1} presented in Theorem 1.
\end{theorem}
\begin{IEEEproof}
See Appendix \ref{appen_Th_2}.
\end{IEEEproof}
In addition, we provide the following corollary to characterize the scaling law of the worst-case CRB on the distance estimation for the optimal APV, i.e., ${\text{CRB}}_r(\mv{x}^\star, r_{\text{max}})$, w.r.t. the array aperture $A$.
\begin{corollary} \label{Cor2}
The associated worst-case CRB on the distance estimation for the optimal APV, i.e., $\text{CRB}_r(\mv{x}^\star, r_{\text{max}})$, decreases with $A$ in the order of ${\cal O}(A^{-4})$. 
\end{corollary}
\begin{IEEEproof}
See Appendix \ref{appen_Cor_2}.
\end{IEEEproof}
It is noteworthy that the optimal APV in Case 1.2 is identical to that in the case of far-field sensing\cite{ma2024movable} as well, since the prior knowledge about the target angle removes the angle-distance coupling. In addition, it is noted from Corollary 2 that the worst-case CRB on the distance estimation for the optimal APV decays faster than that on the AoA ($A^{-4}$ versus $A^{-2}$). This implies that increasing the array aperture sharpens the near-field focal depth more significantly than it narrows the angular main lobe. The reason is that angle estimation depends on the linear phase gradient in \eqref{1D_channel} across the array, whereas distance estimation relies on the quadratic phase variation in \eqref{1D_channel} associated with wavefront curvature.\vspace{-4pt}



\subsection{Joint AoA and Distance Estimation in Case 1.3}
In this subsection, we focus on the joint estimation of the AoA and distance in Case 1.3. Note that the traditional 2D-MUSIC algorithm can leverage the inherent geometric symmetry of the array to decompose the 2D joint estimation problem into two lower-complexity 1D estimation problems \cite{zhang2018localization}. However, MA arrays generally lack such symmetric structure due to their flexible antenna repositioning. Hence, we modify the traditional 2D-MUSIC algorithm by performing an exhaustive search over the 2D angle-distance grids to identify the peaks of the 2D spectrum function \cite{wang2023near}. Therefore, the joint estimation result is given by
\begin{equation}
    \hat{\mv{\eta}} = \arg\max_{\Bar{\mv{\eta}} \in [0, u_{\text{max}}] \times [r_{\text{min}}, r_{\text{max}}]} \frac{1}{\mv{\alpha}(\mv{x}, \Bar{\mv{\eta}})^\mathsf{H} \mv{U}_{\mv{w}} \mv{U}_{\mv{w}}^\mathsf{H} \mv{\alpha}(\mv{x}, \Bar{\mv{\eta}})}.
\label{1_2_2D_search}
\end{equation}
Accordingly, we aim to minimize the CRB on \eqref{1_2_2D_search} by optimizing the MA positions. To this end, we first derive the Fisher information matrix (FIM) of the estimator based on the 2D-MUSIC algorithm. Next, the CRB matrix is derived by taking the inverse of the FIM of the estimator. Specifically, the CRBs on AoA and distance in the joint estimation are given by
\begin{equation}
    \text{CRB}_u(\mv{x}) = \kappa \cdot \frac{\text{var}(\mv{\tilde{x}})}{\text{var}(\mv{x})\text{var}(\mv{\tilde{x}})-{\text{cov}^2(\mv{x},\mv{\tilde{x}})}},
\label{1_3_CRBu}
\end{equation}
\begin{equation}
    \text{CRB}_r(\mv{x}, \mv{\eta}) \!=\! \kappa \cdot \frac{4r^4\text{var}(\mv{x})+8ur^3{\text{cov}(\mv{x},\mv{\tilde{x}})}+4u^2r^2\text{var}(\mv{\tilde{x}})}{(1-u^2)^2\Big(\text{var}(\mv{x})\text{var}(\mv{\tilde{x}})-\text{cov}^2(\mv{x},\mv{\tilde{x}})\Big)},
\label{1_3_CRBr}
\end{equation}
respectively. The procedures for deriving the CRBs are provided in Appendix \ref{appen_1_3}. It can be observed from \eqref{1_3_CRBu} and \eqref{1_3_CRBr} that there may exist a fundamental trade-off between minimizing $\text{CRB}_u(\mv{x})$ and $\text{CRB}_r(\mv{x}, \mv{\eta})$ due to the complicated coupling between the variance and covariance terms therein. Moreover, although $\text{CRB}_u(\mv{x})$ is independent of the AoA, $\text{CRB}_r(\mv{x}, \mv{\eta})$ depends on both the AoA and distance. To overcome this difficulty, we adopt the sum of $\text{CRB}_u(\mv{x})$ and the worst-case $\text{CRB}_r(\mv{x}, \mv{\eta})$, i.e., $\text{CRB}_u(\mv{x})+\max_{\mv{\eta}}\text{CRB}_r(\mv{x}, \mv{\eta})$, as a performance metric to optimize the APV $\mv{x}$. It is noted from \eqref{1_3_CRBr} that $\text{CRB}_r(\mv{x}, \mv{\eta})$ is constantly positive and increases monotonically with both $u$ and $r$. Thus, it reaches the maximum at the maximum values of $u$ and $r$. As such, the estimator vector that yields the worst-case $\text{CRB}_r(\mv{x}, \mv{\eta})$ is given by $\mv{\eta}_{\text{opt}}=[u_{\text{max}},r_{\text{max}}]^\top$. This indicates that, in Case 1.3, the worst-case performance arises when the target lies closest to the \textit{end-fire} direction and farthest from the MA array. The reason is that when the target approaches the end-fire direction, the effective transverse aperture becomes smaller, and the derivatives w.r.t. AoA and distance become more strongly coupled. As a result, the determinant of the FIM decreases, leading to a larger joint-estimation CRB. Moreover, as the target distance increases, the phase curvature becomes weaker, and the wavefront gradually approaches a plane wave. This reduces the Fisher information associated with distance and also results in a larger CRB. The optimization problem in Case 1.3 can be formulated as
\begin{align}
\text{(P1-3)} \; 
& \max_{\mv{x}} \; F_{\mv{\eta}}(\mv{x}) \triangleq \Big(\text{CRB}_u(\mv{x})+\text{CRB}_r(\mv{x}, \mv{\eta}_{\text{opt}})\Big)^{-1} \nonumber\\
& \text{s.t.} \quad \eqref{P1-1b-star}, \eqref{P1-1c-star}. \label{P1-3}
\end{align}
However, it is observed from \eqref{1_3_CRBu} and \eqref{1_3_CRBr} that the objective function of (P1-3) is non-convex w.r.t. the APV $\mv{x}$, rendering (P1-3) challenging to solve optimally. Therefore, we utilize a discrete sampling-based algorithm \cite{mei2024posistion,wei2024joint,mei2024movable} to derive a high-quality sub-optimal APV solution to (P1-3) by sequentially selecting the optimal sampling point for each MA. Note that the sequential update circumvents the difficulty of handling the structural complexity of the CRB objective function. Moreover, compared with exhaustive search, it dramatically reduces computational complexity while ensuring monotonic convergence. In this regard, the sequential update algorithm is also applicable to more general near-field sensing scenarios involving both amplitude and phase curvature, provided that the corresponding CRBs can be derived.

Specifically, the continuous MA array is uniformly discretized into $M$ $(M \gg N)$ sampling points, with the distance between any two adjacent sampling points denoted by $\delta_s =\frac{A}{M}$ and the position of the $i$-th sampling point given by $x_i=i \delta_s, i \in {\cal M} \triangleq \{1, 2, \ldots, M\}$. By denoting ${\cal P}=\{x_i | i \in {\cal M}\}$ as the set of all sampling points, we first construct an initial set of the MA positions, denoted by ${\cal P}_{\text{init}}=\{x^{\text{init}}_n | x^{\text{init}}_n \in {\cal P}, n \in {\cal N}\}$. In the $n$-th iteration, we only update the position of the $n$-th MA, i.e., $x^{\text{init}}_n$, while keeping the positions of the other $(N-1)$ MAs fixed. Let $x^\star_n$ denote the updated position of the $n$-th MA in the $n$-th iteration. Hence, the set of all feasible sampling points for updating $x^{\text{init}}_n$ is
\begin{align}
     {\cal P}_n = \{p | p \in {\cal P}, & \lvert p-x_i^\star \rvert \geq d, 1 \leq i \leq n-1, \lvert p-x^{\text{init}}_j \rvert \geq d, \notag \\
     & n+1 \leq j \leq N\}, 2 \leq n \leq N-1,
\label{pn_set}
\end{align}
and we set ${\cal P}_1 = \{p | p \in {\cal P}, \lvert p-x^{\text{init}}_j \rvert \geq d, 2 \leq j \leq N\}$ and ${\cal P}_N = \{p | p \in {\cal P}, \lvert p-x_i^\star \rvert \geq d, 1 \leq i \leq N-1\}$. Then, we update $x^{\text{init}}_n$ as $x_n^\star$ by minimizing the objective function of (P1-3), i.e.,
\begin{equation}
    x_n^\star = \arg\max_{p \in {\cal P}_n} F_{\mv{\eta}}(\hat{\mv{x}}_{n}),
\label{arg1}
\end{equation}
where $\mv{\hat{x}}_{n}=[x^\star_1,\ldots,x^\star_{n-1},s,x^{\text{init}}_{n+1},\ldots,x^{\text{init}}_N]^\top \in \mathbb{R}^N$. Next, in the $(n+1)$-th iteration, we proceed to update ${\cal P}_{n+1}$ based on \eqref{pn_set} and then update the $(n+1)$-th MA position based on \eqref{arg1}. Note that the above sequential update process can yield a non-decreasing objective function value of (P1-3); hence, its convergence is guaranteed. It is straightforward to see $M - (\frac{2Md}{A} - 1)(N - 1) \leq |{\cal P}_n| \leq M$. As such, the computational complexity order of the proposed sequential update algorithm is between ${\cal O}(MN - N(N - 1)(\frac{2Md}{A} - 1))$ and ${\cal O}(MN)$, and it scales linearly with $M$ for a given $N$. However, it is noteworthy that the algorithm may yield suboptimal solutions because the sets ${\cal P}_n, n \in \mathcal{N}$ are influenced by both the initial selection and the order of sampling points. Additionally, procedures such as Gibbs sampling can be invoked to escape undesired suboptimal solutions \cite{liu2025general}. The main procedures of the proposed algorithm for solving (P1-3) are summarized in Algorithm \ref{algo_case_1_3}. 

\section{MA-Enhanced Near-field sensing for 2D Array}
\label{sec_2D}
\begin{figure}[t]
    \centering
    \includegraphics[scale=0.35]{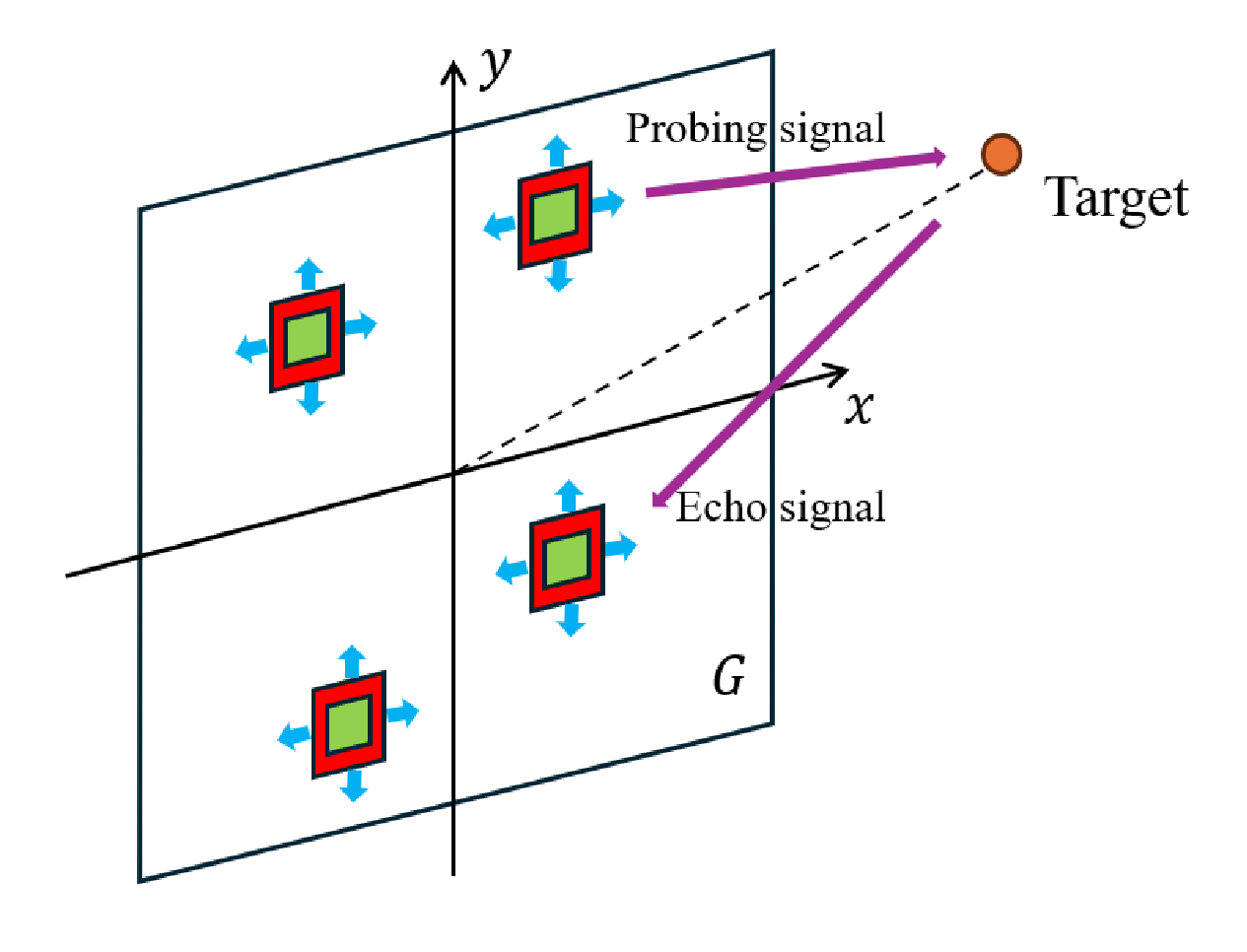}
    \caption{\small 2D MA array for near-field target localization.}
    \label{2D_sysmodel}
    \vspace{-9pt}
\end{figure}
\begin{figure}[t]
    \centering
    \includegraphics[scale=0.4]{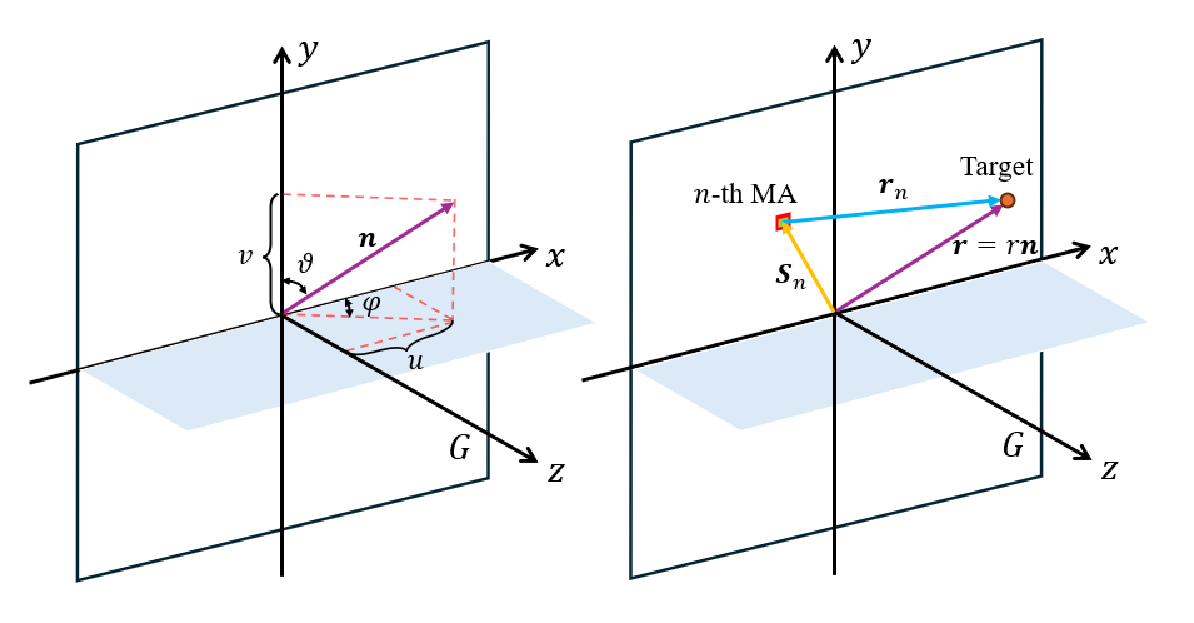}
    \caption{\small Illustration of the target parameters for the 2D MA array.}
    \label{2D_sysmodel_para}
    \vspace{-9pt}
\end{figure}
\subsection{System Model}
In this section, we consider a 2D near-field wireless sensing system with \( N \) MAs to estimate the angular/spatial parameter(s) of a target, as shown in Fig. \ref{2D_sysmodel}. The MAs can move continuously across a 2D square plane denoted by ${\cal G}\triangleq A \times A$, where $A$ is the side length. We assume that the target is located in the radiating near-field region of the planar array, such that the propagating waves have spherical wavefronts. This means that the distance between the target and any position within the MA array is between the Fresnel distance and the Rayleigh distance, which are respectively given by $R_{FS} \triangleq (\frac{A^4}{2\lambda})^{\frac{1}{3}}$ \cite{selvan2017fraunhofer} and $R_{RL} \triangleq \frac{4A^2}{\lambda}$ \cite{liu2023near}.

Let the center of the square plane be the origin and denote the coordinate of the \( n \)-th MA by \( \mv{s}_n \triangleq [x_n, y_n]^\top \in {\cal G}\). Then, the coordinates of the MAs should satisfy $x_n, y_n \in [-\frac{A}{2}, \frac{A}{2}], n \in {\cal N}$. The APM of all \( N \) MAs is denoted by $\tilde{\mv{s}} \triangleq [\mv{s}_1, \mv{s}_2, \ldots, \mv{s}_N] \in \mathbb{R}^{2\times N}$. As depicted in Fig. \ref{2D_sysmodel_para}, the elevation and azimuth steering angles of the LoS path from the origin of the MA array to the target are denoted by $\theta \in [\theta_{\text{min}}, \frac{\pi}{2}]$ and $\phi \in [\phi_{\text{min}}, \frac{\pi}{2}]$, respectively, where $\theta_{\text{min}} > 0$ and $\phi_{\text{min}} > 0$ are the prescribed lower-bounds of the elevation and azimuth AoAs, respectively.\footnote{For analytical simplicity, we only consider the case where the target is in the first quadrant of the plane, i.e., $\theta \in (0, \frac{\pi}{2}]$ and $\phi \in (0, \frac{\pi}{2}]$. In the case that the target is in the other three quadrants, we can flip the MAs along the $x$/$y$-axis and apply our proposed algorithm accordingly.} For analytical convenience, the two AoAs are respectively defined as
\begin{equation}
    u \triangleq \sin \theta \cos \phi \in [0, u_{\text{max}}], \quad v \triangleq \cos \theta \in [0, v_{\text{max}}],
\end{equation}
where $u_{\text{max}}$ and $v_{\text{max}}$ are their upper-bounds, respectively. Then, the unit wave vector of the LoS path can be written as $\mv{n} = [u, v, \sqrt{1-u^2-v^2}]^{\top}$. Denote $r \in [r_{\text{min}}, r_{\text{max}}]$ as the distance between the origin and the target. Hence, the positioning vector of the target is $\mv{r} = r \mv{n}$, and the distance from the \( n \)-th MA to the target can be expressed as a function of the APM $\tilde{\mv{s}}$ and the target parameters, i.e.,
\begin{equation}
\label{Taylor2}
r_n(\mv{s}_n, \mv{\eta}) = \|\mv{r}-\mv{s}_n\| = \sqrt{r^2-2\mv{r}^\top\cdot \mv{s}_n+\|\mv{s}_n\|^2},
\end{equation}
where $\mv{\eta}= [u, v, r]^\top$ is the target parameter vector. By adopting the Fresnel approximation under the near-field model \cite{liu2023near}, the channel coefficient between the \( n \)-th MA and the target is given by
\begin{align}
h_n&(\mv{s}_n, \mv{\eta}) = \\ \notag
&\beta \exp\Big(j\frac{2\pi}{\lambda}\big(x_n u+y_n v - \frac{(x_n^2+y_n^2)-(x_n u+y_n v)^2}{2r}\big)\Big).
\end{align}
Hence, the echoed LoS channel vector can be expressed as
\begin{align}
\mv{h}(\tilde{\mv{s}}, \mv{\eta}) &= [h_1(\mv{s}_1, \mv{\eta}), h_2(\mv{s}_2, \mv{\eta}), \dots, h_N(\mv{s}_N, \mv{\eta})]^\top \\ &= \beta \mv{\alpha}(\tilde{\mv{s}}, \mv{\eta}) \in \mathbb{C}^N, \notag
\end{align}
where $\mv{\alpha}(\tilde{\mv{s}}, \mv{\eta})$ is the near-field steering vector of the 2D MA array. Similar to the 1D scenario presented in Section \ref{sec_1D}, we focus on the following three cases:
\begin{enumerate}
    \item \textbf{Estimation of AoAs only for the 2D MA array (Case 2.1)}: $r$ is known while $u$ and $v$ are both unknown;
    \item \textbf{Estimation of distance only for the 2D MA array (Case 2.2)}: $u$ and $v$ are both known while $r$ is unknown;
    \item \textbf{Joint estimation of AoAs and distance for the 2D MA array (Case 2.3)}: all of $u$, $v$ and $r$ are unknown.
\end{enumerate}

\subsection{AoA Estimation in Case 2.1}
For the AoA estimation in Case 2.1, the distance from the origin of the MA array to the target is assumed to be known and denoted as $r^{\star}$, such that only the two AoAs $u$ and $v$ are estimated. Hence, the estimator vector is $\mv{\eta}=[u, v]^\top$. Similar to the AoA estimation in the 1D MA array case, the estimation of \( u \) and \( v \) via the MUSIC algorithm is given by
\begin{equation}
    \hat{\mv{\eta}} = \arg\max_{\Bar{\mv{\eta}} \in [0, u_{\text{max}}] \times [0, v_{\text{max}}]} \frac{1}{\mv{\alpha}(\tilde{\mv{s}}, \Bar{\mv{\eta}})^\mathsf{H} \mv{U}_{\mv{w}} \mv{U}_{\mv{w}}^\mathsf{H} \mv{\alpha}(\tilde{\mv{s}}, \Bar{\mv{\eta}})},
\end{equation}
which can be obtained by performing an exhaustive search for $\Bar{\mv{\eta}}= [\Bar{u}, \Bar{v}]^\top$ over the interval $[0, u_{\text{max}}] \times [0, v_{\text{max}}]$. Based on the 2D-MUSIC algorithm, the CRB matrix of the estimator vector can be derived. For notational simplicity, we define
\begin{equation}
\xi_n \triangleq x_n+\frac{x_n(x_n u+y_n v)}{r}, n \in {\cal N},
\label{def_xi}
\end{equation}
and
\begin{equation}
\pi_n \triangleq y_n+\frac{y_n(x_n u+y_n v)}{r}, n \in {\cal N}.
\label{def_pi}
\end{equation}
By denoting $\mv{\xi} = [\xi_1, \xi_2, \ldots, \xi_N]^\top \in \mathbb{R}^N$ and $\mv{\pi} = [\pi_1, \pi_2, \ldots, \pi_N]^\top \in \mathbb{R}^N$, the CRBs on the two AoAs for the 2D MA array are respectively given by
\begin{equation}
\text{MSE}(u) \geq \text{CRB}_u(\tilde{\mv{s}}, \mv{\eta}) = \frac{\kappa}{\text{var}(\mv{\xi}) - \frac{\text{cov}^2(\mv{\xi},\mv{\pi})}{\text{var}(\mv{\pi})}} \Bigg|_{r=r^\star},
\label{2_1_CRBu}
\end{equation}
\begin{equation}
\text{MSE}(v) \geq \text{CRB}_v(\tilde{\mv{s}}, \mv{\eta}) = \frac{\kappa}{\text{var}(\mv{\pi}) - \frac{\text{cov}^2(\mv{\xi},\mv{\pi})}{\text{var}(\mv{\xi})}} \Bigg|_{r=r^\star},
\label{2_1_CRBv}
\end{equation}
where the variance functions are defined as $\text{var}(\mv{\xi}) \triangleq \frac{1}{N} \sum_{n=1}^N \xi_n^2 - \mu(\mv{\xi})^2$ with $\mu(\mv{\xi}) = \frac{1}{N} \sum_{n=1}^N \xi_n$ being the mean of \(\mv{\xi}\) and $\text{var}(\mv{\pi}) \triangleq \frac{1}{N} \sum_{n=1}^N \pi_n^2 - \mu(\mv{\pi})^2$ with $\mu(\mv{\pi}) = \frac{1}{N} \sum_{n=1}^N \pi_n$ being the mean of \(\mv{\pi}\). The covariance function is defined as $\text{cov}(\mv{\xi},{\mv{\pi}}) \triangleq \frac{1}{N} \sum_{n=1}^N \xi_n \pi_n - \mu(\mv{\xi})\mu(\mv{\pi})$. The derivations of the above CRBs are provided in Appendix \ref{appen_2_1}.



\begin{figure}[t]
    \centering
    \subfloat[$\theta=86.4^\circ$, $\phi=90^\circ$]{
        \includegraphics[width=0.22\textwidth]{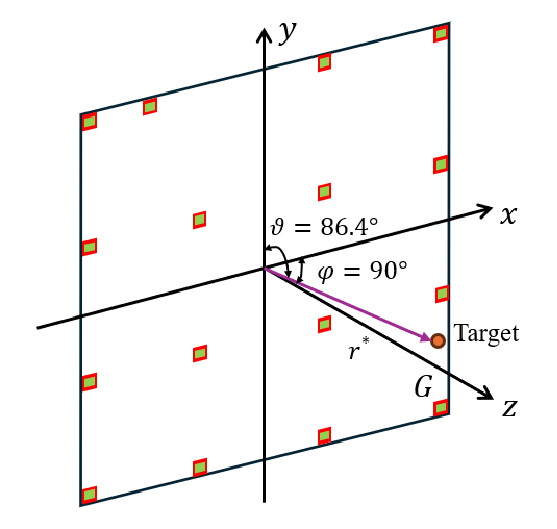}
    }
    \subfloat[$G_{\mv{\eta}}(\tilde{\mv{s}}, \mv{\eta})$]{
        \includegraphics[width=0.25\textwidth]{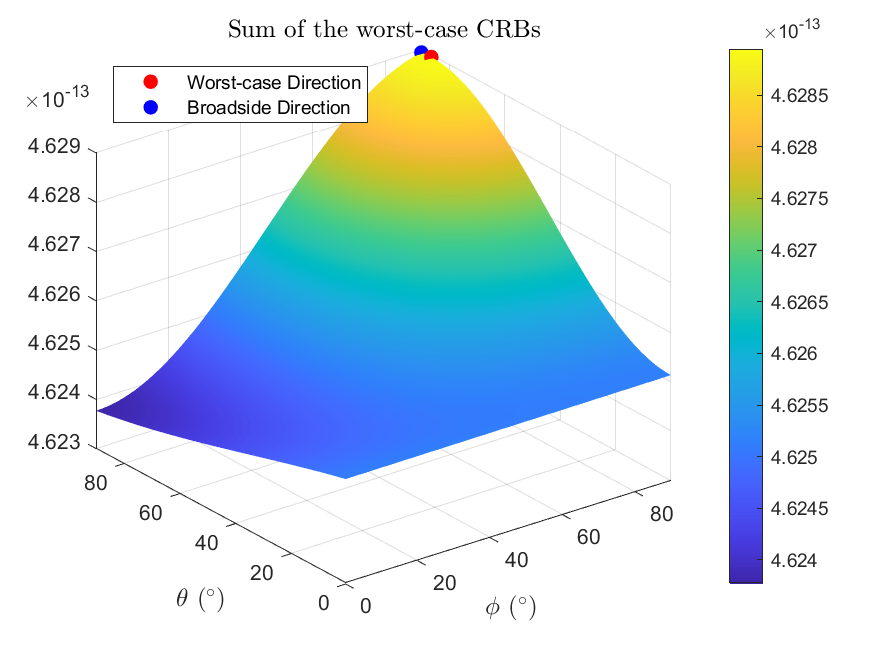}
    }
    \caption{\small Target's elevation and azimuth angles under the worst sensing case for an asymmetric APM.}
    \label{2D_86_90}
    \vspace{-9pt}
\end{figure}

It is noted that the CRBs in \eqref{2_1_CRBu} and \eqref{2_1_CRBv} are not only dependent on the APM but also on the estimator vector $\mv{\eta}$ itself. To overcome this difficulty, we consider minimizing the worst-case sum of the CRBs in \eqref{2_1_CRBu} and \eqref{2_1_CRBv}. To this end, we first denote the target's AoA estimator vector that yields the maximum value of $\text{CRB}_u(\tilde{\mv{s}}, \mv{\eta})+\text{CRB}_v(\tilde{\mv{s}}, \mv{\eta})$ for any given APM $\tilde{\mv{s}}$ as\vspace{-6pt}
\begin{equation}
    \mv{\eta}_{\text{opt}}(\tilde{\mv{s}})=\arg\max_{\bar{\mv{\eta}}\in[0, u_{\text{max}}]\times[0, v_{\text{max}}]} G_{\mv{\eta}}(\tilde{\mv{s}}, \bar{\mv{\eta}}),
\label{eta_opt}
\end{equation}
with $G_{\mv{\eta}}(\tilde{\mv{s}}, \bar{\mv{\eta}}) \triangleq \text{CRB}_u(\tilde{\mv{s}}, \bar{\mv{\eta}})+\text{CRB}_v(\tilde{\mv{s}}, \bar{\mv{\eta}})$. Then, the APM is optimized to minimize $G_{\mv{\eta}}(\tilde{\mv{s}}, \mv{\eta}_{\text{opt}}(\tilde{\mv{s}}))$. However, unlike Cases 1.1–1.3 for 1D MA arrays, where the worst‑case target direction admits a closed‑form solution, it is challenging to derive a closed‑form expression for $\mv{\eta}_{\text{opt}}$ in \eqref{eta_opt}. Consequently, APM optimization for 2D MA arrays incurs considerably higher computational complexity, as it requires a 2D search over the target's AoA to evaluate the performance of each APM.

To circumvent this difficulty, we first consider a symmetric array geometry or APM. Inspired by Case 1.1, we expect that the worst-case performance for AoA estimation for the 2D array also occurs when the target is in the broadside direction, i.e., both its elevation and azimuth angles are $90^\circ$, corresponding to $u=v=0$. Rigorously, for any symmetric APM $\tilde{\mv{s}}$, the results in \eqref{2_1_CRBu} and \eqref{2_1_CRBv} indicate that $G_{\mv{\eta}}(\tilde{\mv{s}}, \mv{\eta})$ achieves the maximum when both $\text{var}(\mv{\xi})$ and $\text{var}(\mv{\pi})$ reach the minimum while $\text{cov}(\mv{\xi},\mv{\pi})$ reaches the maximum. From \eqref{def_xi} and \eqref{def_pi}, it is noted that the variations among the $N$ terms in $\mv{\xi}$ and $\mv{\pi}$ are minimal when $u=v=0$, which yields the minimum $\text{var}(\mv{\xi})$ and $\text{var}(\mv{\pi})$. Moreover, when $u=v=0$, $\mv{\xi}$ and $\mv{\pi}$ degrade to $\mv{x}=[x_1,x_2,\ldots,x_N]^\top$ and $\mv{y}=[y_1,y_2,\ldots,y_N]^\top$, respectively. Since $\mv{x}$ is identical to $\mv{y}$ for any symmetric APM, the linear correlation between the $N$ terms in $\mv{\xi}$ and $\mv{\pi}$ is maximized, which in turn yields the maximum $\text{cov}(\mv{\xi},\mv{\pi})$. It follows from the above that $\mv{\eta}_{\text{opt}}(\tilde{\mv{s}})=[0,0]^\top$ holds for symmetric array geometries. 

However, for an asymmetric array geometry or APM, it is generally difficult to derive the two AoAs that yield the worst-case sum of the CRBs. In fact, for the asymmetric array geometry shown in Fig. \ref{2D_86_90}(a), it can be shown that the worst-case sum of the CRBs is achieved at $\theta = 86.4^\circ$ and $\phi = 90^\circ$ by employing a 2D exhaustive search. Nevertheless, as shown in Fig. \ref{2D_86_90}(b), the difference in $G_{\mv{\eta}}(\tilde{\mv{s}}, \bar{\mv{\eta}})$ between its global maximum and the broadside direction remains below 1\%. In addition to the geometry shown in Fig. \ref{2D_86_90}(a), we have also checked other array geometries and consistently observed small performance differences. The details are omitted due to the page limit. Therefore, we set $\mv{\eta}_{\text{opt}}(\tilde{\mv{s}})=\mv{\eta}_{\text{bd}} \triangleq [0, 0]^\top, \forall \tilde{\mv{s}}$, which greatly simplifies the optimization problem while ensuring both mathematical tractability and minimal performance compromise. In fact, as will be shown in Fig. \ref{2D_Opt_APM}(a) in Section \ref{sec_numerical} via simulation, the optimized array geometry for Case 2.1 is symmetric in both dimensions. 

With $\mv{\eta}_{\text{opt}}(\tilde{\mv{s}}) = [0,0]^{\top}, \forall \tilde{\mv{s}}$, the remaining MA position optimization problem becomes
\begin{subequations}
\label{P2-1}
\begin{align}
\text{(P2-1)} \quad 
& \max_{\tilde{\mv{s}}} \quad F_{\mv{\eta}}(\tilde{\mv{s}}) \triangleq {G^{-1}_{\mv{\eta}}}(\tilde{\mv{s}}, \mv{\eta}_{\text{bd}}) \label{P2-1a} \\
& \text{s.t.} \quad \mv{s}_n \in {\cal G}, \quad n \in {\cal N}, \label{P2-1b} \\
& \phantom{\text{s.t.} \quad} \| \mv{s}_k - \mv{s}_l \| \geq d, \quad k \neq l, \quad k,l \in {\cal N}. \label{P2-1c}
\end{align}
\end{subequations}
Nonetheless, as seen from \eqref{def_xi}, \eqref{def_pi}, \eqref{2_1_CRBu} and \eqref{2_1_CRBv}, the objective function of (P2-1) involves multiple intricate variance and covariance functions, making it highly non-convex w.r.t. the APM $\tilde{\mv{s}}$. Similar to Case 1.3, we utilize the discrete sampling-based algorithm to circumvent this non-convexity and obtain a high-quality sub-optimal APM solution $\tilde{\mv{s}}^\star$ to (P2-1) \cite{mei2024posistion,wei2024joint,mei2024movable}.

\begin{algorithm}[t]
\caption{Proposed Algorithm for Solving (P2-1)}
\label{algo_case_2_1}
\begin{algorithmic}[1]
\State \textbf{Input:} $n=1$, ${\cal Q}$, and ${\cal Q}_{\text{init}}$.
\While{$n \leq N$}
    \State Obtain $\mv{s}^\star_n$ based on \eqref{arg2} and update $\mv{s}^{\text{init}}_n \gets \mv{s}^\star_n$.
    \State Determine ${\cal Q}_{n+1}$ based on \eqref{qn_set}.
    \State Update $n \gets n+1$.
\EndWhile
\State \textbf{Output:} the optimized APM of all $N$ MAs, i.e., $\tilde{\mv{s}}^\star$.
\end{algorithmic}
\end{algorithm}

Specifically, we uniformly sample the continuous 2D MA array into $\tilde{M} \triangleq M^2$ $(\tilde{M} \gg N)$ discrete points, with $M$ points along each dimension. Hence, the inter-point spacing in each dimension is given by $\delta_s = \frac{A}{M}$, and the coordinate of the $(k,l)$-th sampling point is $\mv{s}_{kl}=[k \delta_s, l \delta_s]^\top, k,l\in \bar{{\cal M}} \triangleq \{\frac{-M+1}{2}, \frac{-M+3}{2}, \ldots, \frac{M-3}{2}, \frac{M-1}{2}\}$. The set of all sampling points is denoted as ${\cal Q}=\{\mv{s}_{kl} | k,l \in \bar{{\cal M}}$. Based on this, we initialize the position set as ${\cal Q}_{\text{init}}=\{\mv{s}^{\text{init}}_n | \mv{s}^{\text{init}}_n \in {\cal Q}, n \in {\cal N}\}$. In the $n$-th iteration, the algorithm exclusively adjusts the position of the $n$-th MA, i.e., $\mv{s}^{\text{init}}_n$, while fixing the positions of the remaining $(N-1)$ MAs. Denote $\mv{s}^\star_n$ as the updated position of the $n$-th MA in the $n$-th iteration. Then, the set of all feasible sampling points for updating $\mv{s}^{\text{init}}_n$ is
\begin{align}
     {\cal Q}_n = \{\mv{q} | \mv{q} \in {\cal Q}, & \| \mv{q}-\mv{s}_i^\star \| \geq d, 1 \leq i \leq n-1, \| \mv{q}-\mv{s}^{\text{init}}_j \| \geq d, \notag \\
     & n+1 \leq j \leq N\}, 2 \leq n \leq N-1,
\label{qn_set}
\end{align}
and we set ${\cal Q}_1 = \{\mv{q} | \mv{q} \in {\cal Q}, \| \mv{q}-\mv{s}^{\text{init}}_j \| \geq d, 2 \leq j \leq N\}$ and ${\cal Q}_N = \{\mv{q} | \mv{q} \in {\cal Q}, \| \mv{q}-\mv{s}_i^\star \| \geq d, 1 \leq i \leq N-1\}$. To maximize the objective function of (P2-1), $\mv{s}^{\text{init}}_n$ should be updated as
\begin{equation}
    \mv{s}_n^\star = \arg\max_{\mv{q} \in {\cal Q}_n} F_{\mv{\eta}}(\hat{\mv{s}}_{n}),
\label{arg2}
\end{equation}
where $\hat{\mv{s}}_{n} = [\mv{s}^\star_1,\ldots,\mv{s}^\star_{n-1},\mv{q},\mv{s}^{\text{init}}_{n+1},\ldots,\mv{s}^{\text{init}}_N] \in \mathbb{R}^{2 \times N}$. In the subsequent $(n+1)$-th iteration, we proceed to update ${\cal Q}_{n+1}$ according to \eqref{qn_set} and then derive the position of the $(n+1)$-th MA based on \eqref{arg2}. Importantly, this sequential updating scheme ensures a non-decreasing objective value of (P2-1), thereby guaranteeing convergence. The computational complexity of Algorithm \ref{algo_case_2_1} is between ${\cal O}\big(M^2N - N(N - 1)(\frac{2M^2d}{A} - 1)\big)$ and ${\cal O}(M^2N)$. The key steps of the proposed algorithm are outlined in Algorithm \ref{algo_case_2_1}.\vspace{-3pt}

\subsection{Distance Estimation in Case 2.2}
In this subsection, we focus on the estimation of the target distance under the assumption that the two AoAs are already known and denoted as $u^{\star}$ and $v^{\star}$, respectively. Similar to Case 1.2, the MUSIC algorithm is adopted by utilizing the distance-related information in the signal phase \cite{wang2023near}. By defining
\begin{equation}
\rho_n \triangleq \frac{x_n^2+y_n^2-(x_n u+y_n v)^2}{2r^2}, n \in {\cal N},
\label{def_rho}
\end{equation}
and denoting $\mv{\rho} = [\rho_1, \rho_2, \ldots, \rho_N]^\top \in \mathbb{R}^N$, the CRB on the distance for the 2D MA array is given by
\begin{equation}
    \text{MSE}(r) \geq \text{CRB}_r(\tilde{\mv{s}}, r) = \frac{\kappa}{\text{var}(\mv{\rho})} \Big|_{u=u^\star, v=v^\star}.
\label{2_2_CRBr}
\end{equation}
See Appendix \ref{appen_2_2} for detailed derivations. Similar to Case 1.2, our objective is to minimize the worst-case CRB on the distance among all possible values of distance, i.e., $\max_{r} \text{CRB}_r(\tilde{\mv{s}}, r)$, to eliminate the dependency of the CRB in \eqref{2_2_CRBr} on the distance. Note that the spatial information in the quadratic term of the signal phase, i.e., $x_n u^\star+y_n v^\star+\frac{(x_n^2+y_n^2)(x_n u^\star+y_n v^\star)^2}{2r}$, decays with distance. Hence, the phase variation becomes less significant as the target distance increases, and $r_{\text{max}}$ should yield the worst-case CRB w.r.t. the distance. As such, the corresponding min-max problem can be formulated as
\begin{subequations}
\label{P2-2}
\begin{align}
\text{(P2-2)} \quad
& \max_{\tilde{\mv{s}}} \quad F_r(\tilde{\mv{s}}) \triangleq \text{var}(\mv{\rho}) \Big|_{u=u^\star, v=v^\star, r=r_{\text{max}}} \\
& \text{s.t.} \quad \eqref{P2-1b}, \eqref{P2-1c}. \notag
\end{align}
\end{subequations}
(P2-2) is equivalent to maximizing the variations among the $N$ terms in $\mv{\rho}$ that are correlated with the coordinates of the MAs. Due to the complex objective function, a closed-form optimal solution to (P2-2) cannot be derived similarly as in Case 1.2. As a result, we adopt the discrete-sampling based algorithm again to obtain a high-quality suboptimal solution. The procedures are similar to those of Algorithm \ref{algo_case_2_1} and thus omitted.

\subsection{Joint AoA and Distance Estimation in Case 2.3}
For the joint estimation of the two AoAs and distance in Case 2.3, we employ the 3D-MUSIC algorithm which performs a comprehensive search across the 3D estimator space comprising elevation angle, azimuth angle, and distance to identify the peaks of the spectrum function, i.e.,
\begin{equation}
    \hat{\mv{\eta}} = \arg\max_{\Bar{\mv{\eta}} \in [0, u_{\text{max}}] \times [0, v_{\text{max}}] \times [r_{\text{min}}, r_{\text{max}}]} \frac{1}{\mv{\alpha}(\tilde{\mv{s}}, \Bar{\mv{\eta}})^\mathsf{H} \mv{U}_{\mv{w}} \mv{U}_{\mv{w}}^\mathsf{H} \mv{\alpha}(\tilde{\mv{s}}, \Bar{\mv{\eta}})}.
\end{equation}
Accordingly, the estimation CRBs on the joint estimation of the two AoAs and distance are respectively given by \eqref{2_3_CRBu}, \eqref{2_3_CRBv}, and \eqref{2_3_CRBr} at the top of this page. The details are provided in Appendix \ref{appen_2_3}.
\begin{figure*}[t]
\begin{equation}
\small
    \widetilde{\text{CRB}}_u(\tilde{\mv{s}}, \mv{\eta}) = \frac{\kappa \big(\text{var}(\mv{\pi})\text{var}(\mv{\rho})-\text{cov}^2(\mv{\pi},\mv{\rho})\big)}{\text{var}(\mv{\xi})\text{var}(\mv{\pi})\text{var}(\mv{\rho})+2\text{cov}(\mv{\xi},\mv{\pi})\text{cov}(\mv{\xi},\mv{\rho})\text{cov}(\mv{\pi},\mv{\rho})-\text{var}(\mv{\xi})\text{cov}^2(\mv{\pi},\mv{\rho})-\text{var}(\mv{\pi})\text{cov}^2(\mv{\xi},\mv{\rho})-\text{var}(\mv{\rho})\text{cov}^2(\mv{\xi},\mv{\pi})},
\label{2_3_CRBu}
\end{equation}
\begin{equation}
\small
    \widetilde{\text{CRB}}_v(\tilde{\mv{s}}, \mv{\eta}) = \frac{\kappa \big(\text{var}(\mv{\xi})\text{var}(\mv{\rho})-\text{cov}^2(\mv{\xi},\mv{\rho})\big)}{\text{var}(\mv{\xi})\text{var}(\mv{\pi})\text{var}(\mv{\rho})+2\text{cov}(\mv{\xi},\mv{\pi})\text{cov}(\mv{\xi},\mv{\rho})\text{cov}(\mv{\pi},\mv{\rho})-\text{var}(\mv{\xi})\text{cov}^2(\mv{\pi},\mv{\rho})-\text{var}(\mv{\pi})\text{cov}^2(\mv{\xi},\mv{\rho})-\text{var}(\mv{\rho})\text{cov}^2(\mv{\xi},\mv{\pi})},
\label{2_3_CRBv}
\end{equation}
\begin{equation}
\small
    \widetilde{\text{CRB}}_r(\tilde{\mv{s}}, \mv{\eta}) = \frac{\kappa \big(\text{var}(\mv{\xi})\text{var}(\mv{\pi})-\text{cov}^2(\mv{\xi},\mv{\pi})\big)}{\text{var}(\mv{\xi})\text{var}(\mv{\pi})\text{var}(\mv{\rho})+2\text{cov}(\mv{\xi},\mv{\pi})\text{cov}(\mv{\xi},\mv{\rho})\text{cov}(\mv{\pi},\mv{\rho})-\text{var}(\mv{\xi})\text{cov}^2(\mv{\pi},\mv{\rho})-\text{var}(\mv{\pi})\text{cov}^2(\mv{\xi},\mv{\rho})-\text{var}(\mv{\rho})\text{cov}^2(\mv{\xi},\mv{\pi})}.
\label{2_3_CRBr}
\end{equation}
{\noindent} \rule[-10pt]{18cm}{0.05em}
\end{figure*}

Similar to Case 1.3, we aim to minimize the worst-case sum of the above three CRBs w.r.t. the estimators. First, we derive the estimator vector $\mv{\eta}_{\text{opt}}(\tilde{\mv{s}})$ that leads to the worst-case sensing performance for any given APM $\tilde{\mv{s}}$. Inspired by Case 1.3, we expect that the worst sensing performance occurs when the target is in the closest direction to end-fire and farthest from the MA array, i.e., $u=0, v=v_{\text{max}}, r=r_{\text{max}}$ or $u=u_{\text{max}},v=0, r=r_{\text{max}}$. To corroborate this, we tested a large number of symmetric and asymmetric array geometries through extensive simulations. The results consistently show only marginal performance gaps between the true worst-case estimator and the above estimator, while a rigorous mathematical proof is left for future work. Hence, we set $\mv{\eta}_{\text{opt}}(\tilde{\mv{s}}) = \mv{\eta}_{\text{ed}} \triangleq [0,v_{\text{max}},r_{\text{max}}]^\top$ in this paper. Note that setting $\mv{\eta}_{\text{opt}}(\tilde{\mv{s}}) = [u_{\text{max}},0,r_{\text{max}}]^\top$ yields the same sensing result by simply rotating the MA array by $90^\circ$ about the $z$-axis. The remaining APM optimization problem becomes
\begin{subequations}
\label{P2-3}
\begin{align}
\text{(P2-3)} \; 
\max_{\tilde{\mv{s}}} \; \tilde{F}_{\mv{\eta}}(\tilde{\mv{s}}) & \triangleq \Big(\widetilde{\text{CRB}}_u(\tilde{\mv{s}}, \mv{\eta}_{\text{ed}})+\widetilde{\text{CRB}}_v(\tilde{\mv{s}}, \mv{\eta}_{\text{ed}})\nonumber\\
&\quad+\widetilde{\text{CRB}}_r(\tilde{\mv{s}}, \mv{\eta}_{\text{ed}})\Big)^{-1} \label{P2-3a} \\
\text{s.t.} \quad & \eqref{P2-1b}, \eqref{P2-1c}, \notag
\end{align}
\end{subequations}
where $\widetilde{\text{CRB}}_u(\tilde{\mv{s}}, \mv{\eta}_{\text{ed}})$, $\widetilde{\text{CRB}}_v(\tilde{\mv{s}}, \mv{\eta}_{\text{ed}})$ and $\widetilde{\text{CRB}}_r(\tilde{\mv{s}}, \mv{\eta}_{\text{ed}})$ denote the CRB on the two AoAs and distance, respectively. It is observed that the objective function of (P2-3) contains even more complex terms w.r.t. the APM $\tilde{\mv{s}}$ compared with (P2-1) and (P2-2), as seen from \eqref{2_3_CRBu}, \eqref{2_3_CRBv} and \eqref{2_3_CRBr}. As such, the discrete sampling-based algorithm is adopted again, with the main procedures the same as Algorithm \ref{algo_case_2_1}.

\section{Numerical Results}\label{sec_numerical}
In this section, numerical results are presented to evaluate the worst-case CRB performance of the proposed MA-enhanced near-field sensing scheme. Unless otherwise stated, the simulation parameters are as follows. The carrier frequency is set to 15 GHz so that the wavelength is $\lambda=0.02$ m. The minimum separation between adjacent MAs is set to $d = \lambda/2$. The average received signal-to-noise ratio (SNR) is defined as $P \lvert\beta\rvert^2 / \sigma^2$. For the 1D MA array, we set the ground-truth target AoAs in Cases 1.1 and 1.3 and the known target AoA in Case 1.2 as $\theta=45^\circ$, i.e., $u =\cos\theta=0.71$; and we set the ground-truth target distances in Cases 1.2 and 1.3 and the known target distance in Case 1.1 as $r=R_{RL}/4$. For the 2D MA array, we set the ground-truth target AoAs in Cases 2.1 and 2.3 and the known target AoA in Case 2.2 as $\theta=\phi=45^\circ$, i.e., $u=\sin\theta \cos\phi=0.50$ and $v=\cos\theta=0.71$; and we set the ground-truth target distances in Cases 2.2 and 2.3 and the known target distance in Case 2.1 as $r=R_{RL}/4$. Moreover, in Cases 1.2, 1.3, 2.2, and 2.3, we set $r_{\text{min}} = R_{FS}$ and $r_{\text{max}} = R_{RL}/2$, i.e., $r \in [R_{FS}, R_{RL}/2]$. In Cases 1.1 and 1.3, we set $\theta_{\text{min}} = \pi/10$ with $u_{\text{max}} = 0.95$; while in Cases 2.1 and 2.3, we set $\theta_{\text{min}} = \phi_{\text{min}} = \pi/10$ with $u_{\text{max}} = v_{\text{max}} = 0.95$. In Algorithms \ref{algo_case_1_3} and \ref{algo_case_2_1}, we set the number of sampling points as $M=10(N-1)+1$. Furthermore, we include the genetic algorithm (GA) as an optimization baseline for comparison. Denote the population size, the maximum number of generations, and the number of independent runs in the GA as $N_{\rm pop}$, $N_{\rm gen}$, and $N_{\rm run}$, respectively. Then, the overall computational complexity of the GA is $\mathcal{O}(N_{\rm run}N_{\rm gen}N_{\rm pop}N^2)$, which is substantially higher than that of the sequential update algorithm since $N_{\rm run}N_{\rm gen}N_{\rm pop} \gg M^2$ usually holds. In the simulation, we set $N_{\rm pop}=N_{\rm gen}=1000$ and $N_{\rm run}=10$.

\subsection{1D MA Array}
We first consider the 1D MA array with $N = 20$ antennas and $A = 20\lambda$. For performance comparison, we consider the following benchmarks:
\begin{enumerate}
    \item \textbf{Benchmark 1.1: Uniform linear array (ULA) with half-wavelength inter-antenna spacing}: $\{x_n\}_{n=1}^N$ are set as $x_n=(n-1)d,n \in {\cal N}$;
    \item \textbf{Benchmark 1.2: Sparse ULA with a full aperture}: $\{x_n\}_{n=1}^N$ are set as $x_n=(n-1)A/(N-1),n \in {\cal N}$.
    \item \textbf{Benchmark 1.3: Optimal MA array in the far-field AoA estimation}: $\{x_n\}_{n=1}^N$ are set the same as those provided in Theorem \ref{Th1}.
    \item \textbf{Benchmark 1.4: Optimized MA array based on GA}: $\{x_n\}_{n=1}^N$ are obtained by adopting the GA.
\end{enumerate}

\begin{figure}[t]
    \centering
    \subfloat[Case 1.1]{
        \includegraphics[width=0.245\textwidth]{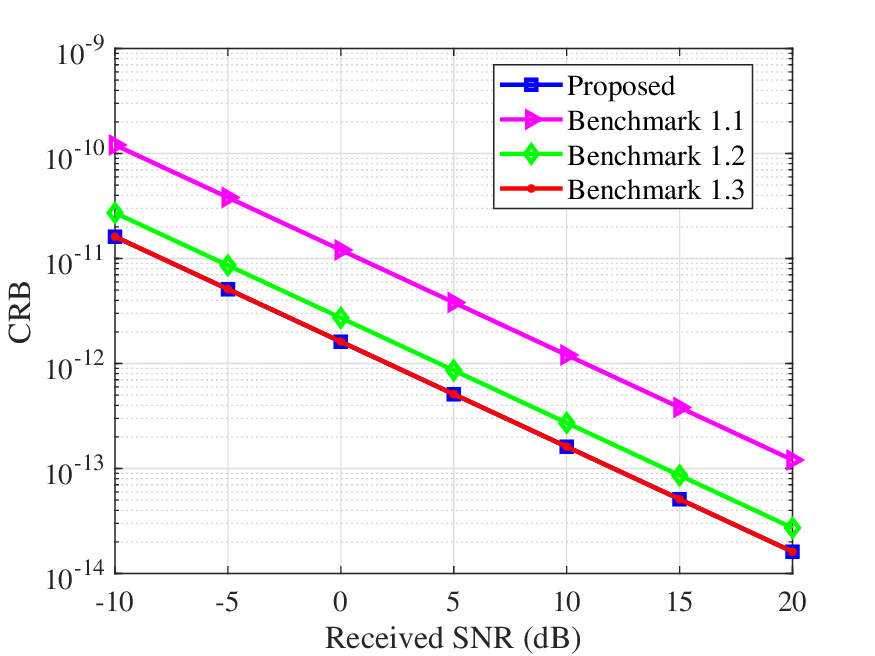}
    }
    \subfloat[Case 1.2]{
        \includegraphics[width=0.245\textwidth]{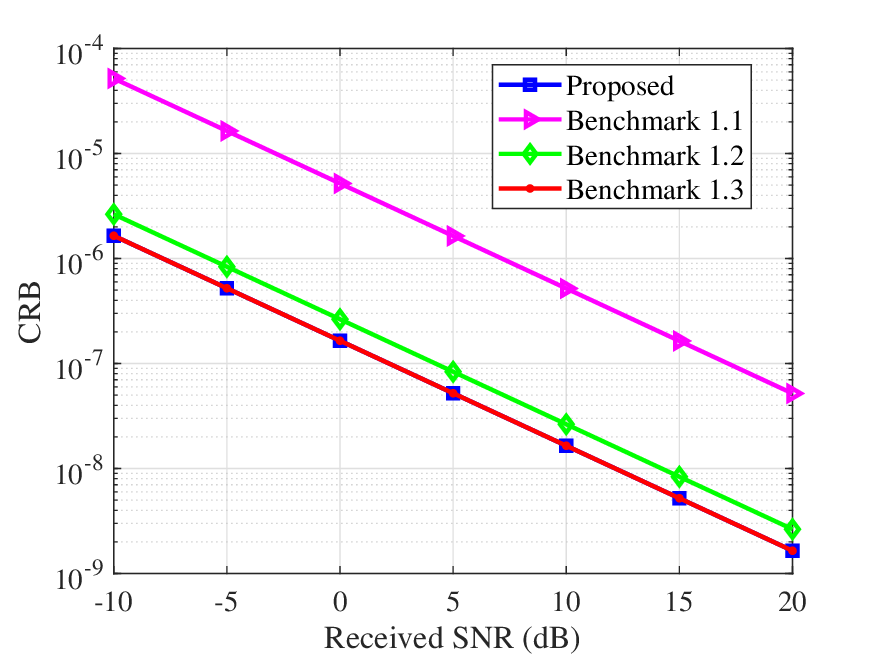}
    }
    \caption{\small Estimation CRBs versus the received SNR for the 1D array in Cases 1.1 and 1.2.}
    \label{Sim_1D_CRB}
\vspace{-9pt}
\end{figure}
In Fig. \ref{Sim_1D_CRB}(a), we show the worst-case CRBs of the AoA estimation MSEs in \eqref{1_1_CRBu} versus the received SNR by different schemes. Note that we do not present the MUSIC MSE in the simulation, as it corresponds to a specific target location and therefore offers no meaningful comparison with the worst‑case CRB, which is evaluated over the entire target region. It is observed that the proposed optimal APV in Theorem \ref{Th1} results in a significantly lower CRB compared with Benchmarks 1.1 and 1.2. For SNR = 20 dB, the optimal APV is observed to yield a 55.3\% and a 20.5\% CRB reduction over Benchmarks 1.1 and 1.2, respectively. Benchmark 1.1 is observed to achieve the worst performance among all considered schemes, as its effective aperture is the smallest, resulting in limited angular resolution. The proposed scheme and Benchmark 1.3 achieve identical performance, which is consistent with Theorem 1. For the distance estimation, the worst-case CRBs in \eqref{1_2_CRBr} versus the received SNR by different schemes are shown in Fig. \ref{Sim_1D_CRB}(b). Similar observations made from Fig. \ref{Sim_1D_CRB}(a) can also be made in Fig. \ref{Sim_1D_CRB}(b). Particularly, compared with Benchmarks 1.1 and 1.2, the proposed scheme leads to a notable decrease in the CRB. For SNR = 20 dB, the CRB is reduced by 74.2\% and 18.4\% over Benchmarks 1.1 and 1.2, respectively. Benchmark 1.3 achieves an identical performance to the proposed scheme again, which is consistent with Theorem 2.

\begin{figure}[t]
    \centering
    \subfloat[Optimized APV]{
        \includegraphics[width=0.48\textwidth]{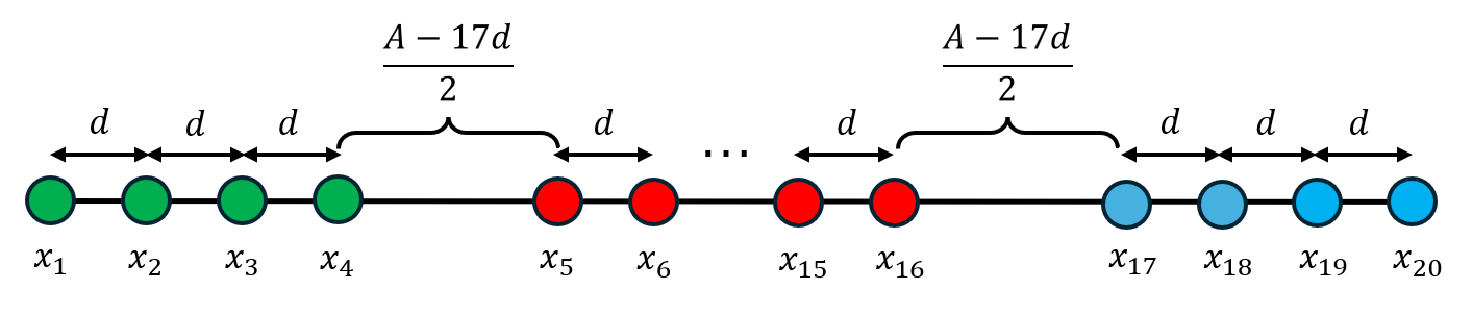}
    }
    \\
    \vspace{-6pt}
    \subfloat[Case 1.3]{
        \includegraphics[width=0.36\textwidth]{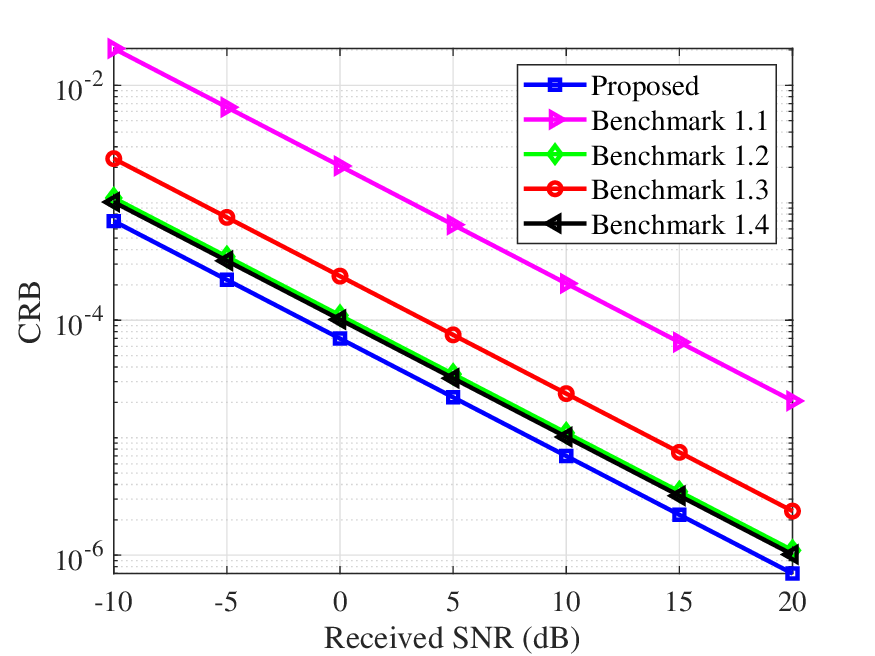}
    }
    \caption{\small Optimized positions of MAs and estimation CRBs versus the received SNR for the 1D array in Case 1.3.}
    \label{1D_MAarray_Case3}
    \vspace{-9pt}
\end{figure}
For the joint estimation of AoA and distance, we obtain the optimized MA array geometry by adopting Algorithm \ref{algo_case_1_3} and initializing the antenna geometry as a sparse ULA with a full aperture, i.e., Benchmark 1.2. In Fig. \ref{1D_MAarray_Case3}(a), we show the optimized positions of the MAs for the joint estimation of AoA and distance. It is observed that, unlike the array geometry shown in Fig. \ref{1D_MAarray_Case1_2} for AoA/distance estimation only, that for joint estimation consists of three groups of antennas, as marked by different colors. This is because placing all MAs at the boundaries makes the linear term $\mv{x}$ and the quadratic term $\mv{\tilde{x}}$ highly linearly correlated, which causes the determinant of the FIM $\text{var}(\mv{x})\text{var}(\mv{\tilde{x}})-{\text{cov}^2(\mv{x},\mv{\tilde{x}})}$ to approach zero, thus causing singularity in the joint estimation. Therefore, placing some MAs in the central region is necessary to decorrelate the angle and distance information. Moreover, the first and the $N$-th MAs are placed at the two endpoints of the array, respectively, i.e., $x_1=0$ and $x_N=A$, which maximizes the array aperture to increase estimation resolution. To verify the effectiveness of the proposed scheme, we plot the worst-case sums of the CRBs in \eqref{1_3_CRBu} and \eqref{1_3_CRBr} versus the received SNR in Fig. \ref{1D_MAarray_Case3}(b). It is observed that the optimized APV by the sequential update algorithm results in a remarkable decrease in the CRB compared with Benchmarks 1.1-1.4. Specifically, for SNR = 20 dB, the proposed scheme achieves a 96.6\%, 36.5\%, 70.6\% and 31.1\% reduction over Benchmarks 1.1-1.4, respectively. Particularly, Benchmark 1.3 is observed to incur a significant performance loss in joint angle and distance estimation. This is because the far-field MA geometry mainly exploits AoA-related phase curvature information and may not adequately capture distance‑related phase curvature.

\begin{figure}[t]
    \centering
    \includegraphics[scale=0.42]{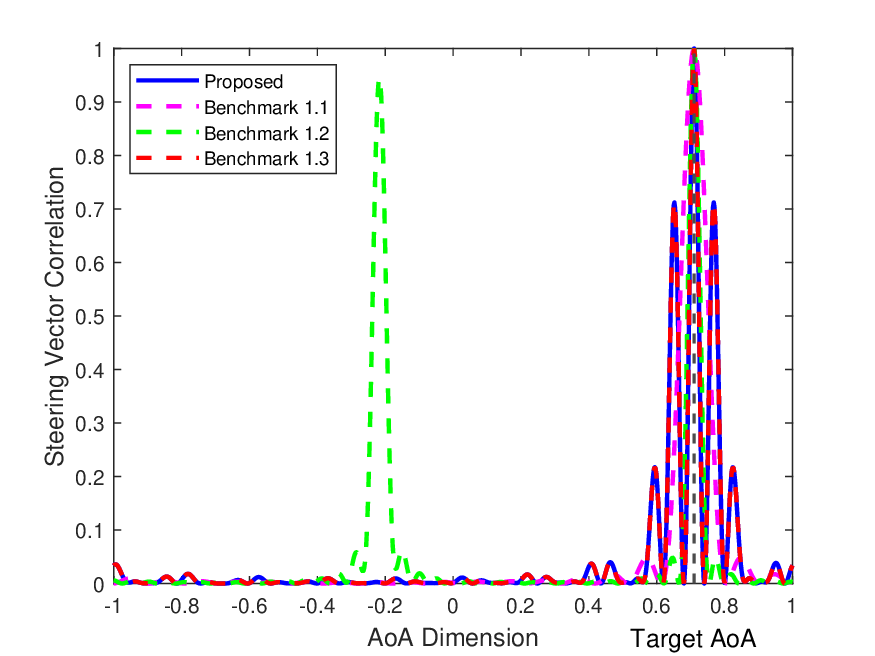}
    \caption{\small Steering vector correlation in Case 1.1.}
    \label{1_1_steer_corr}
    \vspace{-9pt}
\end{figure}


\begin{figure}[t]
    \centering
    \includegraphics[scale=0.42]{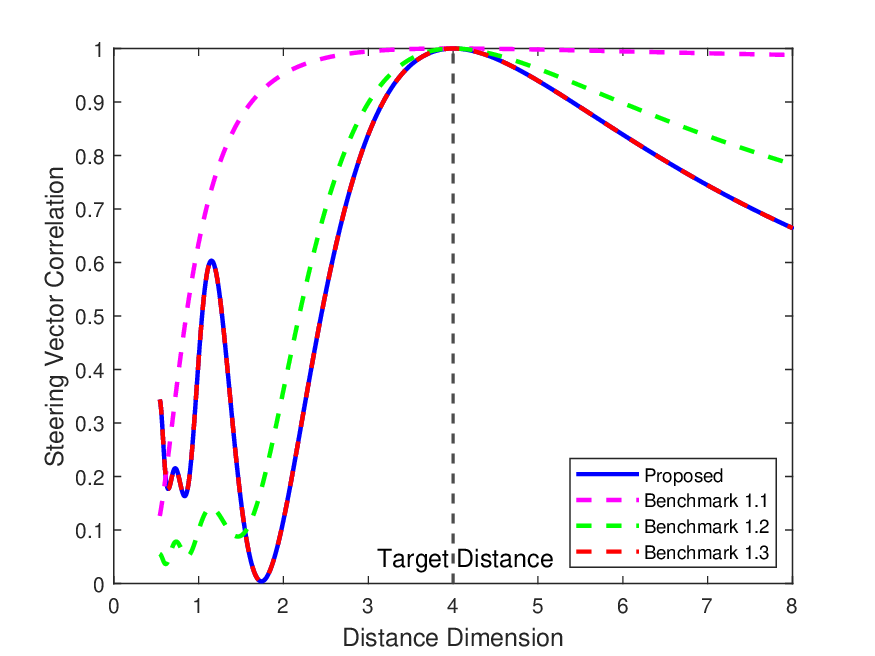}
    \caption{\small Steering vector correlation in Case 1.2.}
    \label{1_2_steer_corr}
    \vspace{-9pt}
\end{figure}


\begin{figure*}[t]
    \centering
    \subfloat[Proposed]{
        \includegraphics[width=0.245\textwidth]{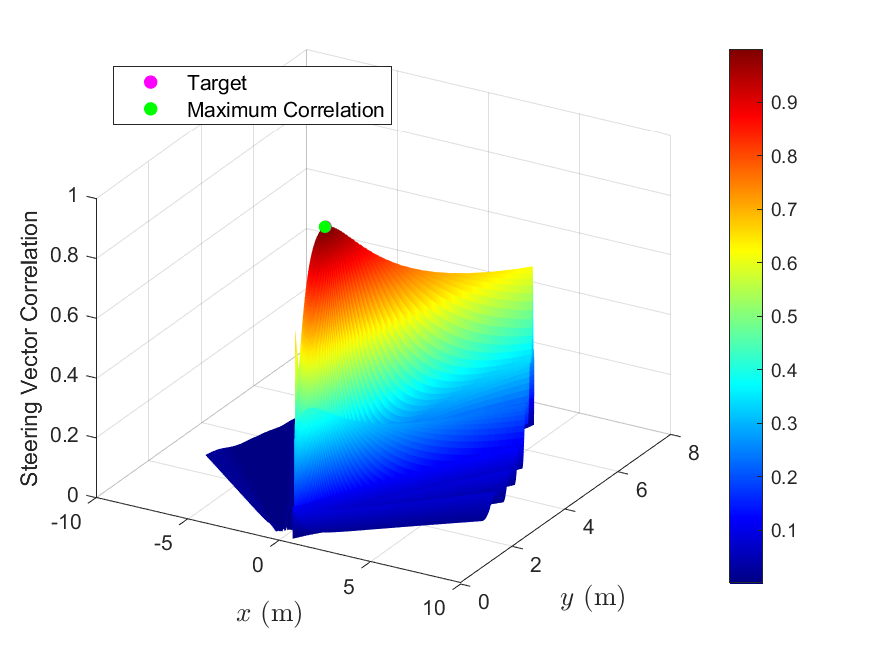}
    }
    \subfloat[Benchmark 1.1]{
        \includegraphics[width=0.245\textwidth]{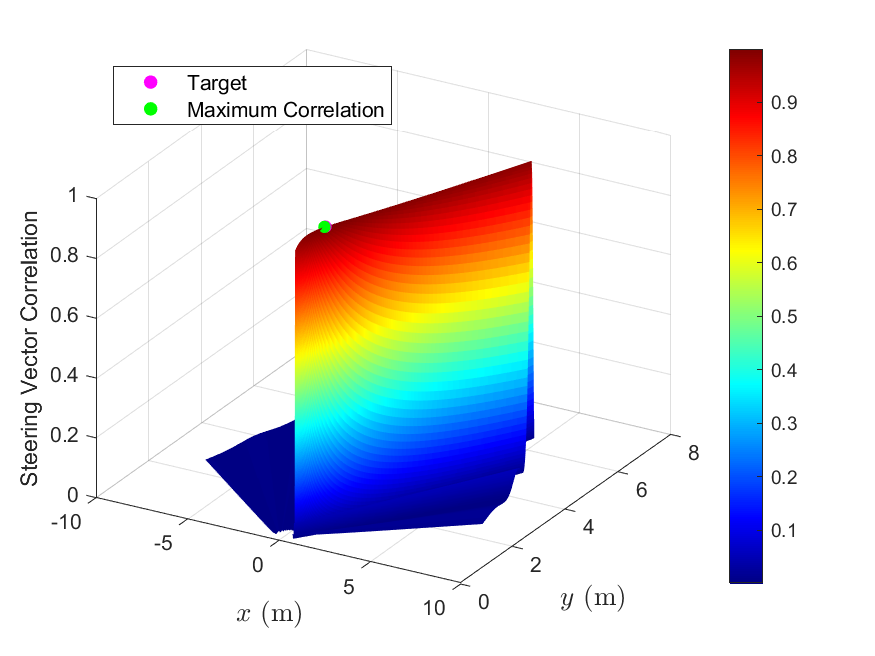}
    }
    \subfloat[Benchmark 1.2]{
        \includegraphics[width=0.245\textwidth]{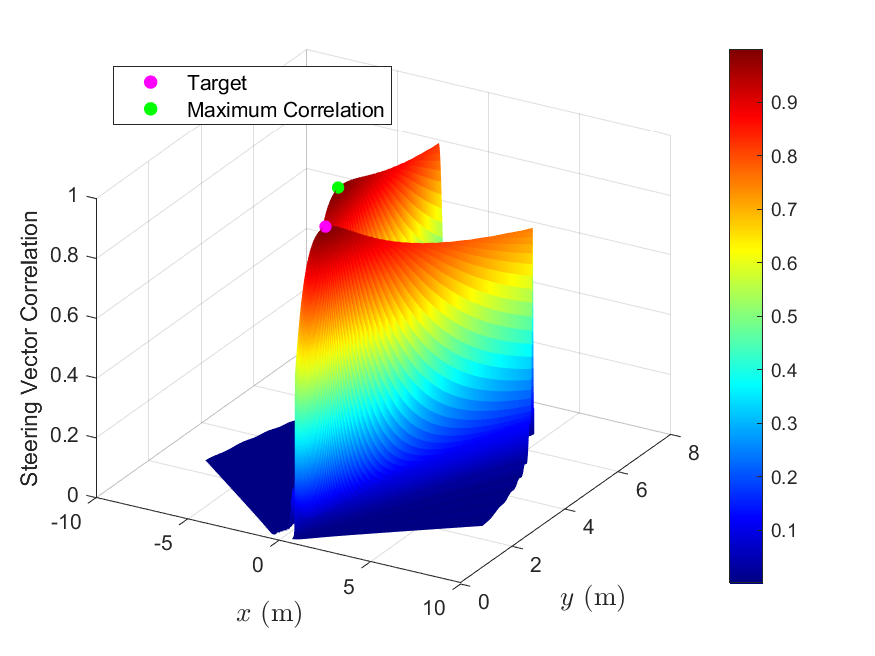}
    }
    \centering
    \subfloat[Benchmark 1.3]{
        \includegraphics[width=0.245\textwidth]{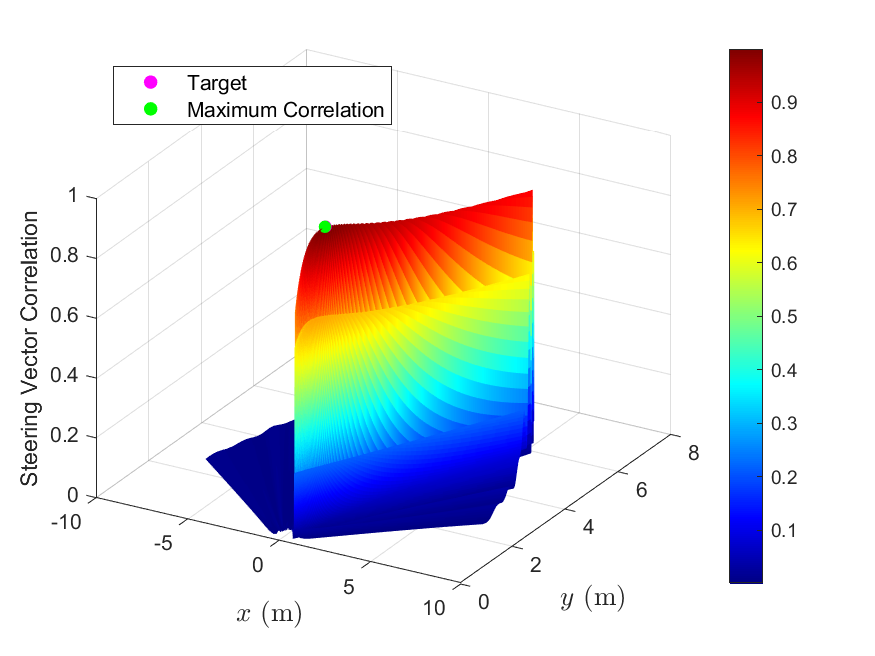}
    }
    \caption{\small Steering vector correlation in Case 1.3.}
    \label{1_3_steer_corr}
    \vspace{-9pt}
\end{figure*}

To reveal more insights, we show the steering vector correlation under the considered schemes in Figs. \ref{1_1_steer_corr}, \ref{1_2_steer_corr} and \ref{1_3_steer_corr}, corresponding to Cases 1.1, 1.2, and 1.3, respectively. The steering vector correlation is defined as $R({\mv{\eta}}') \triangleq \frac{1}{N^2}|\mv{\alpha}(\mv{x}',\mv{\eta})^\mathsf{H}\mv{\alpha}(\mv{x}',{\mv{\eta}}')|^2$, where ${\mv{\eta}}'$ denotes any feasible target parameter (vector), and $\mv{x}'$ denotes the optimized APV by any considered scheme. Evidently, it is desirable for the correlation function $R({\mv{\eta}}')$ to approach a Dirac function, i.e.,
$\mv{R}({\mv{\eta}}')\rightarrow 
\begin{cases}
1, & {\mv{\eta}}' = \mv{\eta}, \\
0, & {\mv{\eta}}' \neq \mv{\eta},
\end{cases}$
thereby achieving higher angular and spatial resolution while reducing ambiguity in parameter estimation. As observed from Figs. \ref{1_1_steer_corr} and \ref{1_2_steer_corr}, the proposed scheme provides a narrower main lobe compared with Benchmarks 1.1 and 1.2. Moreover, it is observed from Fig. \ref{1_1_steer_corr} that a grating lobe occurs at $-0.23$ in Benchmark 1.2. This results in ambiguity in distinguishing the actual AoA $0.71$ from its false estimate at $-0.23$, leading to a large CRB for Benchmark 1.2. Additionally, by comparing Figs. \ref{1_1_steer_corr} and  \ref{1_2_steer_corr}, it is observed that the main lobe in the distance estimation is much broader than that in the AoA estimation, which indicates higher sensing accuracy in the angular domain compared with the distance domain. This is also revealed in Figs. \ref{Sim_1D_CRB}(a) and \ref{Sim_1D_CRB}(b), where the CRBs on the AoA estimation are significantly lower than those on the distance estimation. For the joint estimation, we project $R({\mv{\eta}}')$ onto the $x$-$y$ domain with $(x,y) = (ru, r\sqrt{1-u^2})$ and present the results for all five schemes in Figs. \ref{1_3_steer_corr}(a)-(e). It is observed from Fig. \ref{1_3_steer_corr}(a) that the correlation function for the proposed scheme reaches its maximum at the location of the target with suppressed sidelobes in the angular domain, indicating enhanced resolution and reduced ambiguity. In contrast, Benchmark 1.1 yields a wider main lobe along the distance domain, thereby limiting its spatial resolution and leading to a larger CRB, as shown in Fig. \ref{1_3_steer_corr}(b). For Benchmark 1.2, Fig. \ref{1_3_steer_corr}(c) illustrates an additional undesired ridge-shaped pattern caused by the grating lobe at $-0.23$, which creates AoA estimation ambiguity and leads to a larger CRB compared with the proposed scheme. Fig. \ref{1_3_steer_corr}(d) further shows that the far-field MA geometry in Benchmark 1.3 yields a beam pattern that is wider than that in the proposed scheme but narrower than that in Benchmark 1.1 along the distance domain.\vspace{-6pt}


\subsection{2D MA Array}

\begin{figure*}[t]
    \centering
    \subfloat[Case 2.1]{
        \includegraphics[width=0.32\textwidth]{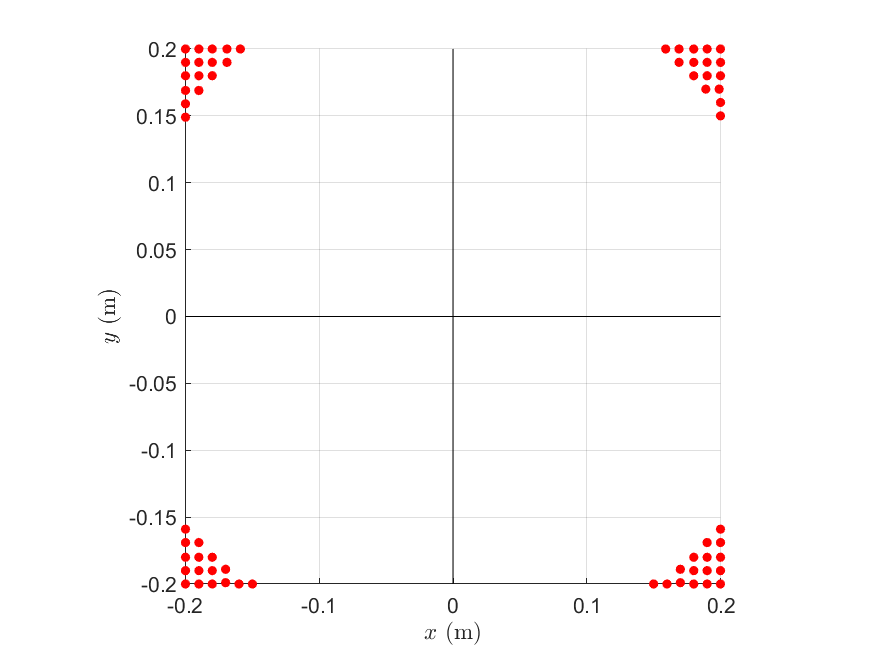}
    }
    \subfloat[Case 2.2]{
        \includegraphics[width=0.32\textwidth]{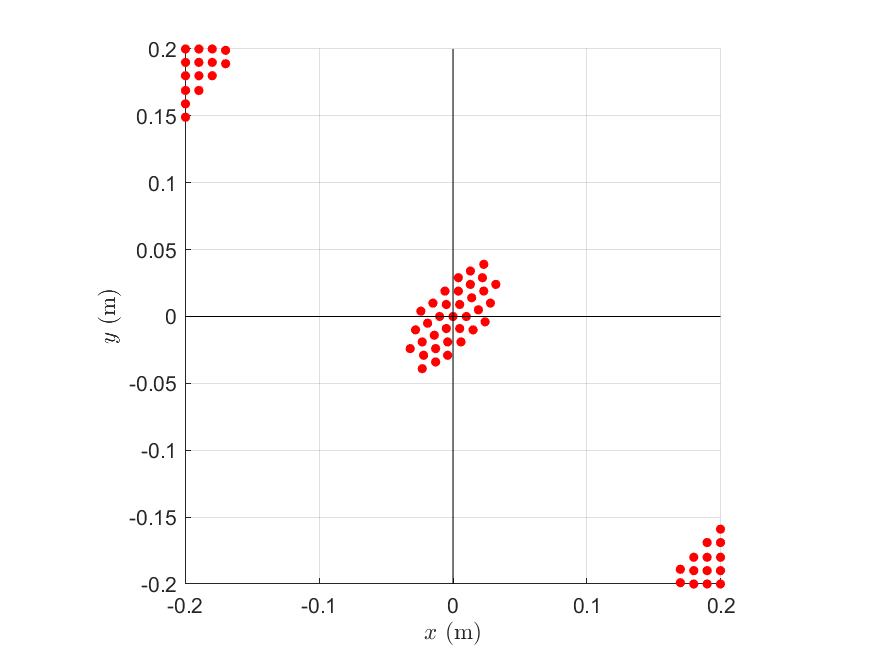}
    }
    \subfloat[Case 2.3]{
        \includegraphics[width=0.32\textwidth]{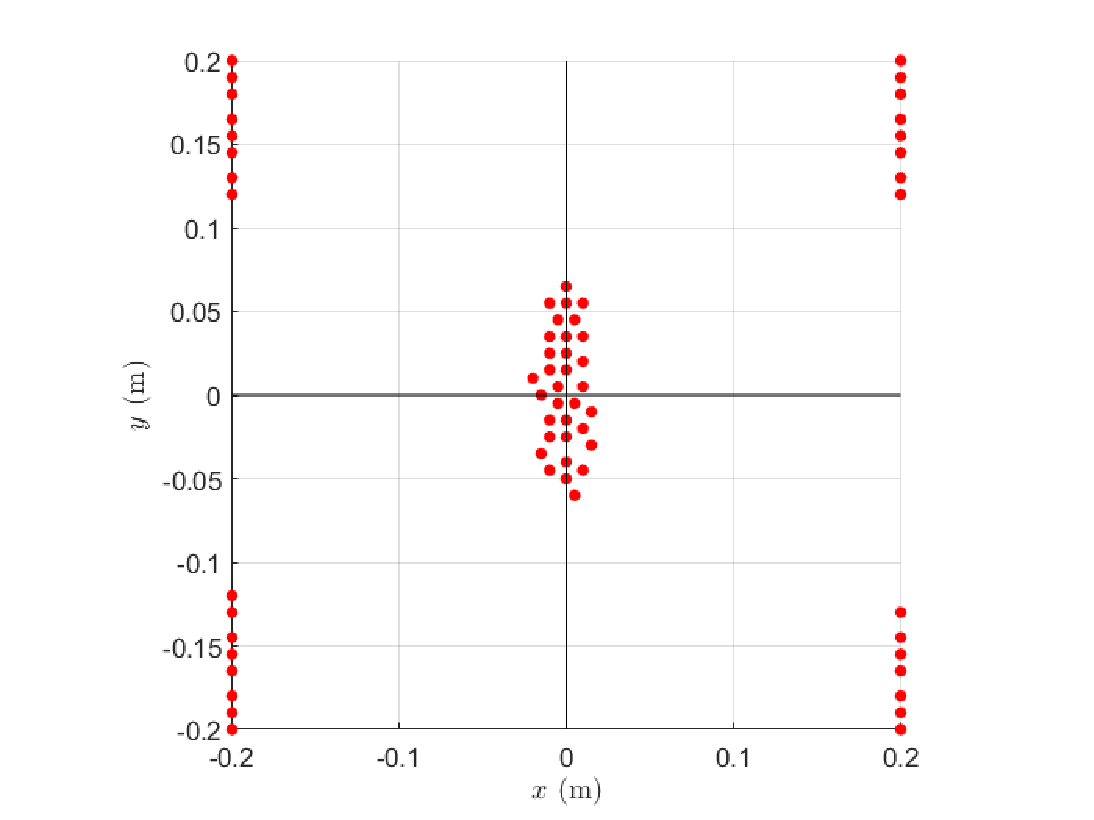}
    }
    \caption{\small Optimized positions of MAs for the 2D MA array by the proposed method.}
    \label{2D_Opt_APM}
\end{figure*}


\begin{figure*}[t]
    \centering
    \subfloat[Case 2.1]{
        \includegraphics[width=0.32\textwidth]{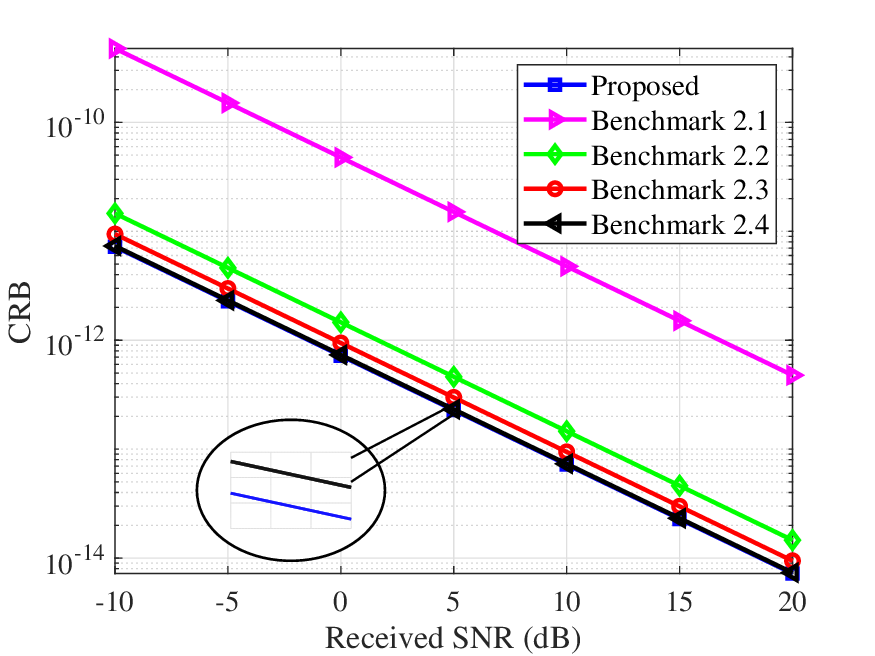}
    }
    \subfloat[Case 2.2]{
        \includegraphics[width=0.32\textwidth]{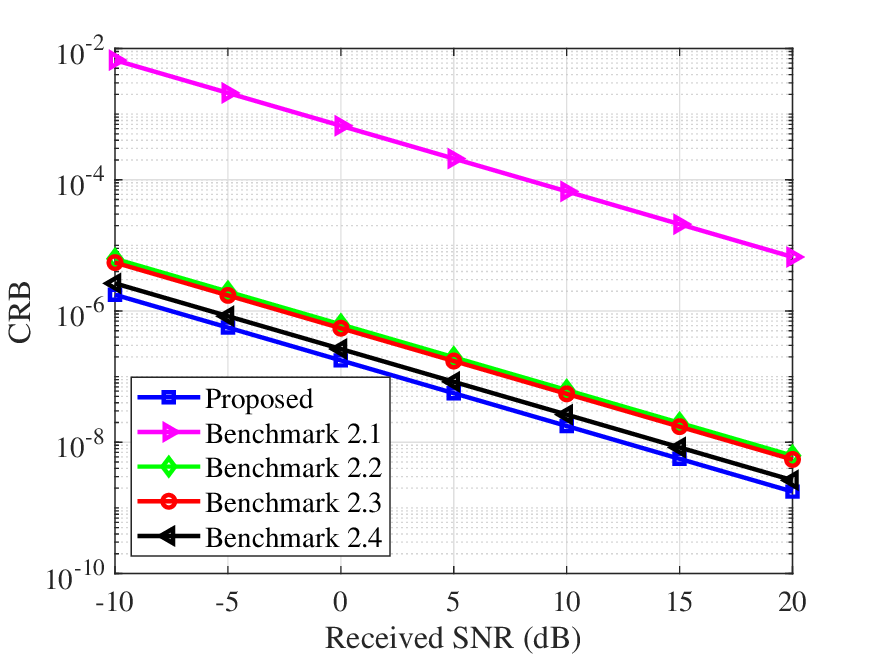}
    }
    \subfloat[Case 2.3]{
        \includegraphics[width=0.32\textwidth]{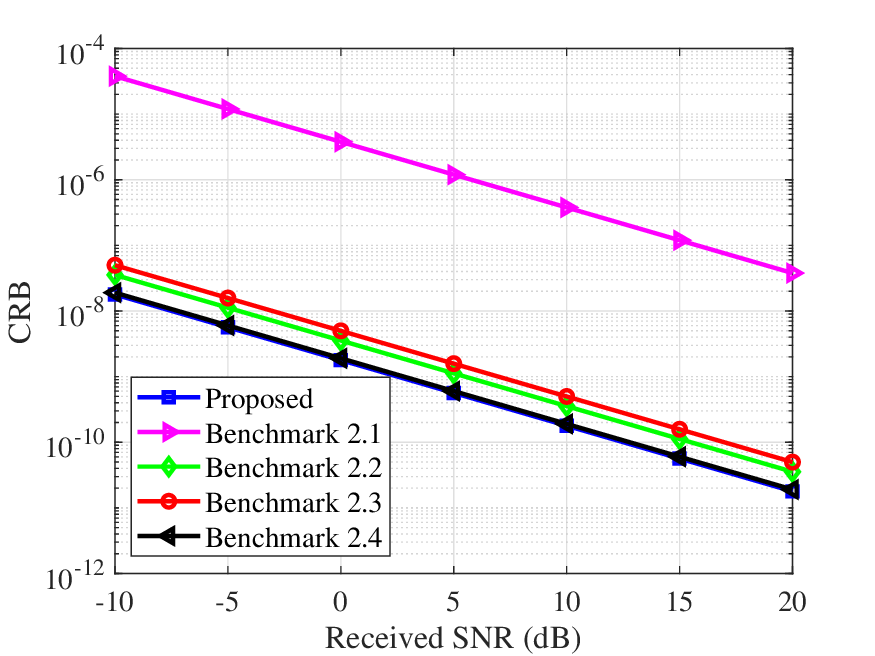}
    }
    \caption{\small Estimation CRBs versus the received SNR for the 2D MA array.}
    \label{Sim_2D_CRB}
    \vspace{-9pt}
\end{figure*}


\begin{figure*}[t]
    \centering
    \subfloat[Target distance of 3 m]{
        \includegraphics[width=0.32\textwidth]{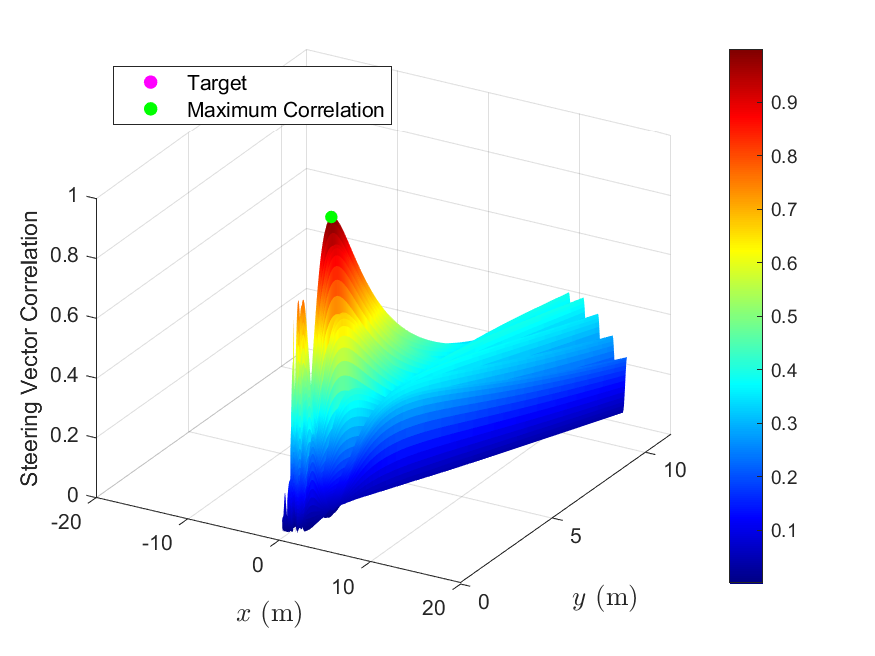}
    }
    \subfloat[Target distance of 8 m]{
        \includegraphics[width=0.32\textwidth]{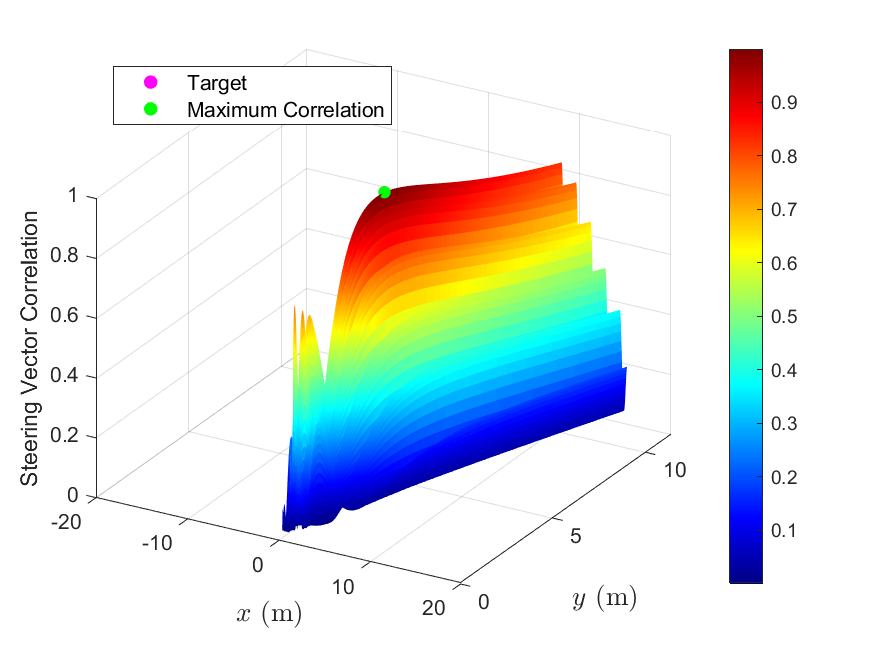}
    }
    \subfloat[Target distance of 15 m]{
        \includegraphics[width=0.32\textwidth]{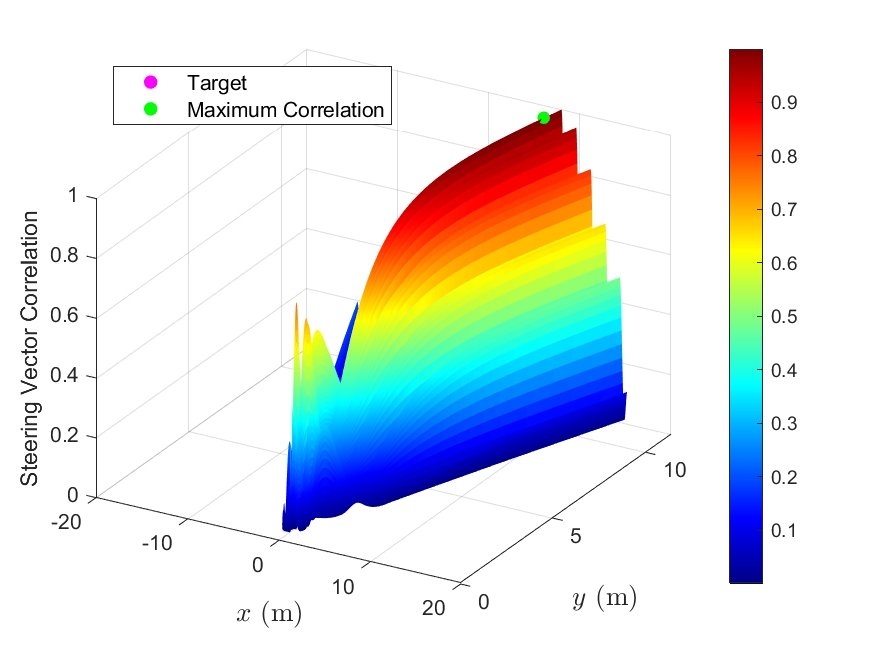}
    }
    \caption{\small Steering vector correlation of different target distances between the target and the 2D MA array.}
    \label{NFS_vs_FFS}
    \vspace{-9pt}
\end{figure*}

In this subsection, we consider the 2D MA array with $N = 8\times 8$ antennas and side length $A = 20\lambda$. The following benchmarks are considered for comparison:
\begin{enumerate}
    \item \textbf{Benchmark 2.1: UPA with half-wavelength antenna spacing}: $\tilde{\mv{s}}$ is set as a UPA with half-wavelength inter-antenna spacing in each dimension;
    \item \textbf{Benchmark 2.2: Sparse UPA with a full aperture}: $\tilde{\mv{s}}$ is set as a UPA with the largest aperture $A \times A$, with the inter-antenna spacing $A/(\sqrt{N}-1)$ in each dimension.
    \item \textbf{Benchmark 2.3: Optimal MA array in the far-field AoA estimation}: $\tilde{\mv{s}}$ is optimized as the APM for far-field AoA estimation based on the alternating optimization (AO) algorithm proposed in \cite{ma2024movable}.
    \item \textbf{Benchmark 2.4: Optimized MA array based on GA}: $\tilde{\mv{s}}$ is obtained by adopting the GA.
\end{enumerate}

Note that we initialize the APM $\tilde{\mv{s}}$ as that in Benchmark 2.2 in Algorithm \ref{algo_case_2_1}. The optimized MA positions for Cases 2.1, 2.2 and 2.3 by adopting Algorithm \ref{algo_case_2_1} are shown in Figs. \ref{2D_Opt_APM}(a)-(c), respectively. It is observed that there always exist MAs deployed on the corners of the square plane in the proposed scheme, thereby achieving the maximum aperture to improve the sensing resolution. Particularly,  Figs. \ref{2D_Opt_APM}(a) and (c) show that the array geometries in Cases 2.1 and 2.3 exhibit near symmetry along both dimensions, which achieves a balance between the estimation of $u$ and $v$, as evidenced in \cite{ma2024movable}. Moreover, the optimized array geometry in Case 2.3 combines the key characteristics of both AoA and distance estimation, as shown in Fig. \ref{2D_Opt_APM}(c). It preserves the MAs at the four corners in Case 2.1 and also the MAs around the central region in Case 2.2, thereby capturing the advantages of both configurations. Figs. \ref{Sim_2D_CRB}(a)-(c) show the worst-case (sums) of the CRBs by different schemes versus the received SNR in Cases 2.1, 2.2, and 2.3, respectively. It is observed that the proposed scheme achieves significantly lower CRBs compared with Benchmarks 2.1 and 2.2. In addition, unlike the 1D MA array, Benchmark 2.3 is no longer competitive with the proposed scheme even in Cases 2.1 and 2.2, due to the more intricate coupling between angle and distance in 2D MA arrays. Benchmark 2.4 is observed to achieve comparable performance to the proposed scheme at the cost of much higher complexity.

Next, we examine the steering vector correlation for the optimized 2D MA array, i.e., $R({\mv{\eta}}')$, in Case 2.3, by projecting it onto the $x$-$y$ domain with $(x,y) = (ru,rv)$. With $v=0.71$, Figs. \ref{NFS_vs_FFS}(a)-(c) show $R({\mv{\eta}}')$ over the $x$-$y$ domain for the target distances of 3 m, 8 m and 15 m (all between $r_{\text{min}}=0.54$ m and $r_{\text{max}}=16.00$ m). Similar correlation patterns as those shown in Figs. \ref{1_1_steer_corr}, \ref{1_2_steer_corr} and \ref{1_3_steer_corr} are observed. Particularly, Fig. \ref{NFS_vs_FFS}(a) reveals a sharply peaked main lobe at the true target location, highlighting the strong \textit{beamfocusing} capability of the optimized array geometry for accurate target localization. However, as the target distance increases to 8 m and then to 15 m, undesired sidelobes along the AoA dimension gradually intensify, while the main lobe along the distance dimension becomes noticeably flatter, reducing its directivity and consequently degrading localization accuracy. This behavior is expected because larger target distances cause the spherical wavefronts to gradually transition towards planar ones, thereby weakening near-field beamfocusing. Nevertheless, Fig. \ref{NFS_vs_FFS} also shows that the main lobe along the AoA direction remains sharply peaked even at large distances. This indicates that near-field target sensing remains largely comparable to its far-field counterpart in terms of AoA estimation.

\begin{figure}[t]
    \centering
    \includegraphics[scale=0.42]{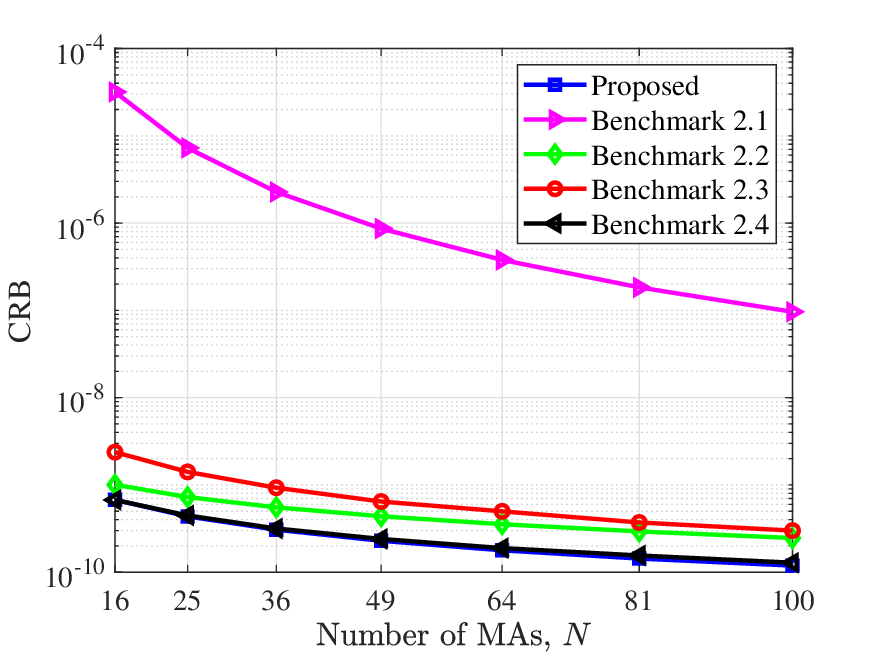}
    \caption{\small Estimation CRBs versus number of MAs in Case 2.3.}
    \label{2_3_CRB_N}
    \vspace{-9pt}
\end{figure}
\begin{figure}[t]
    \centering
    \includegraphics[scale=0.42]{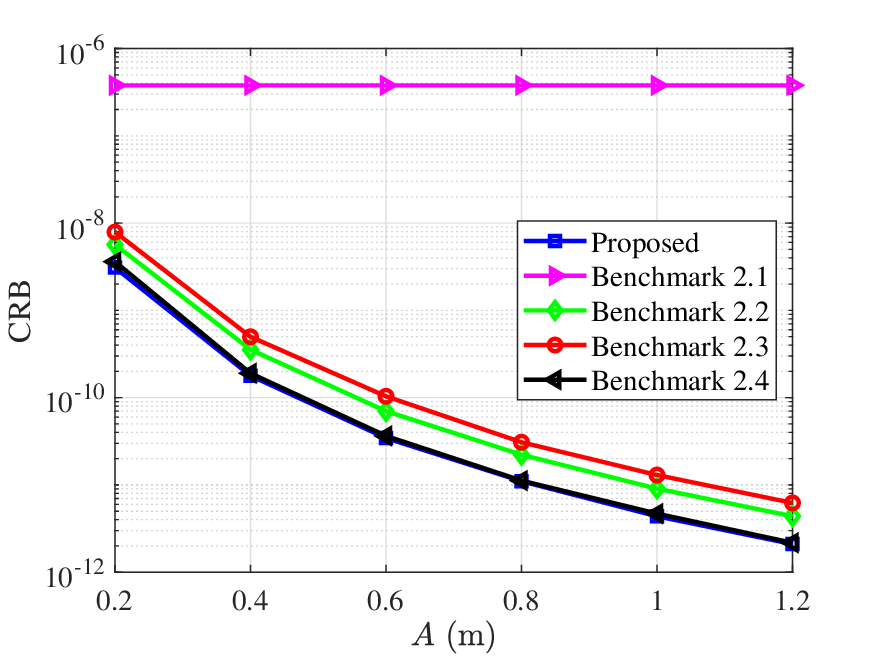}
    \caption{\small Estimation CRBs versus side length of the movement region in Case 2.3.}
    \label{2_3_CRB_A}
    \vspace{-9pt}
\end{figure}



Moreover, in Figs. \ref{2_3_CRB_N} and \ref{2_3_CRB_A}, we compare the worst-case sums of the CRBs by different schemes in Case 2.3 versus the number of MAs $N$ and the side length of the movement region $A$, respectively. The received SNR is set to 10 dB. It is observed that the worst-case sums of the CRBs decrease with either $N$ or $A$, except for Benchmark 2.1 in Fig. \ref{2_3_CRB_A}, which remains unchanged with $A$, as its array geometry does not vary with the array aperture. In addition, Fig. \ref{2_3_CRB_N} shows that the proposed 2D MA array with only $N=16$ MAs achieves a 33.0\%, 71.7\%, and 99.3\% reduction in the CRB compared with Benchmarks 2.2 and 2.3 with $N=16$ MAs and Benchmark 2.1 with even $N=100$ MAs, respectively. In Fig. \ref{2_3_CRB_A}, even for a small-scale array with a side length of $A=0.2$ m, the proposed scheme is observed to yield a 45.5\%, 60.9\%, and 99.2\% decrease in the CRB compared with Benchmarks 2.2, 2.3, and 2.1, respectively. These observations suggest that MAs offer a cost-effective solution for near-field target localization, as they can significantly reduce the number of required antennas compared with FPAs while even achieving superior localization accuracy.

\section{Conclusion}
\label{sec_conclusion}
In this paper, we investigated a new MA-enhanced near-field sensing system employing 1D and 2D antenna arrays, aiming to estimate a target’s AoA and/or distance information using the MUSIC algorithm. We first derived the worst-case (sum) CRBs on the MSEs for both individual and joint estimation of the target’s AoA and distance, and then minimized them via antenna position optimization. For the 1D MA array, closed-form solutions to the individual AoA and distance CRB minimization problems were obtained, revealing that the optimal array geometry is consistent with that of far-field target sensing. To solve the CRB minimization problems in the remaining cases, a low-complexity discrete sampling-based algorithm was proposed to sequentially update the MA positions. Numerical results demonstrated that MAs can substantially reduce the CRBs even with far fewer antennas compared with FPAs by forming narrower main lobes towards the target direction while suppressing grating lobes elsewhere. Furthermore, for joint angle and distance estimation, the optimal MA array geometry was shown to differ from that of individual estimation. This paper can be extended to various directions as future work. First, this paper only considers single-target near-field sensing under the LoS-dominated channel with the second-order Fresnel approximation. More general cases with multiple targets and/or an exact spherical-wave channel model deserve further study. In addition, it would be interesting to study 3D MA arrays for near-field sensing, which provide more spatial DoFs but pose more challenging CRB characterization and optimization problems.

\appendices
\section{Derivations of the CRB in Case 1.1} \label{appen_1_1}
Based on \cite[Theorem 4.1]{stoica1989music}, we can obtain the Fisher information of $u$ for estimating the AoA of the target as\footnote{Note that unbiasedness and regularity of the estimators in all considered cases can be readily demonstrated. Therefore, the CRBs on the estimation MSEs for both individual and joint estimation exist.}
\begin{multline}
    J_u(\mv{x}, u) = \frac{2}{\sigma^2}\sum_{t=1}^{T}\Re\Bigg\{s_t^* \mv{\psi}(\mv{x}, u)^\mathsf{H} \bigg(\mv{I}_N-\mv{h}(\mv{x}, u) \\
    \Big(\mv{h}(\mv{x}, u)^\mathsf{H} \mv{h}(\mv{x}, u)\Big)^{-1}\mv{h}(\mv{x}, u)^\mathsf{H}\bigg)\mv{\psi}(\mv{x}, u)s_t\Bigg\} \Bigg|_{r=r^\star},
\label{app_a_FI}
\end{multline}
where
\begin{equation}
    \mv{\psi}_u(\mv{x}, u) = \frac{\partial \mv{h}(\mv{x}, u)}{\partial u} = j\frac{2\pi}{\lambda}\mv{\zeta}_u \odot \mv{h}(\mv{x}, u)
\label{1_1_psi}
\end{equation} denotes the partial derivative of the near-field channel vector $\mv{h}(\mv{x}, u)$ w.r.t. the estimator $u$, where $\mv{\zeta}_u \triangleq [\zeta_{u,1},\zeta_{u,2},\ldots,\zeta_{u,N}]^\top \in \mathbb{R}^N$ with $\zeta_{u,n} \triangleq x_n+\frac{x_n^2 u}{r}, n \in {\cal N}$. By re-denoting $J_u(\mv{x}, u)$, $\mv{\psi}_u(\mv{x}, u)$ and $\mv{h}(\mv{x}, u)$ as $J_u$, $\mv{\psi}$ and $\mv{h}$ respectively for brevity, \eqref{app_a_FI} can be further expressed as
\begin{align}
    J_u &= \frac{2}{\sigma^2}\sum_{t=1}^{T}\Re\Bigg\{s_t^* \mv{\psi}^\mathsf{H} \bigg(\mv{I}_N-\frac{1}{N}\mv{h} \mv{h}^\mathsf{H}\bigg)\mv{\psi}s_t\Bigg\}\Bigg|_{r=r^\star} \notag \\
    &= \frac{2TP}{\sigma^2} \Re\Bigg\{\mv{\psi}^\mathsf{H}\mv{\psi}- \frac{1}{N\lvert\beta\rvert^2}\mv{\psi}^\mathsf{H}\mv{h}(\mv{\psi}^\mathsf{H}\mv{h})^*\Bigg\}\Bigg|_{r=r^\star} \notag \\ 
    &= \frac{2TP}{\sigma^2} \Re\Bigg\{(\frac{2\pi}{\lambda})^2 \lvert\beta\rvert^2 \sum_{n=1}^{N}\zeta_{u,n}^2-\frac{1}{N\lvert\beta\rvert^2}\Big(-j\frac{2\pi}{\lambda} \lvert\beta\rvert^2 \notag \\
    &\qquad \sum_{n=1}^{N}\zeta_{u,n}\Big)\Big(j\frac{2\pi}{\lambda} \lvert\beta\rvert^2 \sum_{n=1}^{N}\zeta_{u,n}\Big)\Bigg\}\Bigg|_{r=r^\star} \notag \\
    &= \frac{8\pi^2TPN\lvert\beta\rvert^2}{\sigma^2 \lambda^2} \text{var}(\mv{\zeta}_u)\Bigg|_{r=r^\star} \notag \\
    &= \frac{8\pi^2TPN\lvert\beta\rvert^2}{\sigma^2 \lambda^2} \left(\text{var}(\mv{x})+\frac{2u}{r^\star}\text{cov}(\mv{x},\mv{\tilde{x}})+\frac{u^2}{{r^\star}^2}\text{var}(\mv{\tilde{x}}) \right) \notag \\
    &= \frac{F_u(\mv{x}, u)}{\kappa}.
\label{app_a_FI_extend}
\end{align}
Then, the CRB on the AoA estimation for the 1D MA array is obtained by taking the inverse of the Fisher information, i.e.,
\begin{equation}
    \text{CRB}_{u}(\mv{x}, u) = J_u^{-1}(\mv{x}, u) = \frac{\kappa}{F_u(\mv{x}, u)}.
\label{1_1_reciprocal}
\end{equation}
This thus completes the derivations.\vspace{-9pt}

\section{Derivations of the CRB in Case 1.2} \label{appen_1_2}
In the distance estimation for the 1D MA array, the partial derivative of the near-field channel vector $\mv{h}(\mv{x}, r)$ w.r.t. the estimator $r$ is
\begin{equation}
    \mv{\psi}_r(\mv{x}, r) = \frac{\partial \mv{\alpha}(\mv{x}, r)}{\partial r} = j\frac{2\pi}{\lambda}\mv{\zeta}_r \odot \mv{\alpha}(\mv{x},r),
\label{1_2_psi}
\end{equation}
where $\mv{\zeta}_r \triangleq [\zeta_{r,1},\zeta_{r,2},\ldots,\zeta_{r,N}]^\top \in \mathbb{R}^N$ with $\zeta_{r,n} \triangleq \frac{x_n^2(1-u^2)}{2r^2}, n \in {\cal N}$. Hence, the Fisher information of $r$ for estimating the distance of the target is given by
\begin{align}
    J_r(\mv{x}, r) &= \frac{8\pi^2TPN\lvert\beta\rvert^2}{\sigma^2 \lambda^2} \text{var}(\mv{\zeta}_r)\Bigg|_{u=u^\star} \notag \\
    &= \frac{8\pi^2TPN\lvert\beta\rvert^2}{\sigma^2 \lambda^2} \Big(\frac{1-{u^\star}^2}{2r^2}\Big)^2 \text{var}(\tilde{\mv{x}}).
\label{app_b_FI}
\end{align}
As a result, the CRB on the distance estimation is given by
\begin{equation}
    \text{CRB}_{r}(\mv{x}, r) = J_r^{-1}(\mv{x}, r) = \frac{\kappa}{F_r(\mv{x}, r)}.
\end{equation}
This thus completes the derivations.
\vspace{-9pt}

\section{Proof of Theorem \ref{Th2}} \label{appen_Th_2}
For any given APV $\mv{x}$ that fulfills constraints \eqref{P1-1b-star} and \eqref{P1-1c-star}, we demonstrate that $\text{var}(\tilde{\mv{x}}^\star) \geq \text{var}(\tilde{\mv{x}})$ by iteratively optimizing the MA positions $\{x_n\}_{n=1}^N$ through the following sequential adjustment process.

The antenna position adjustment includes $N$ sequential steps, each adjusting the position of one MA while keeping those of the remaining $N-1$ MAs fixed. Define $\mv{x}^{(i-1)} \triangleq [x_1^{(i-1)}, x_2^{(i-1)}, \ldots, x_N^{(i-1)}]^\top$ as the APV prior to the $i$-th adjustment ($i = 1, 2, \ldots, N$) with $\mv{x}^{(0)} = \mv{x}$ and $\tilde{\mv{x}}^{(i-1)} \triangleq \mv{x}^{(i-1)} \odot\mv{x}^{(i-1)}$. In the $i$-th ($i = 1, 2, \ldots, \lfloor N/2 \rfloor$) iteration, the position of the $i$-th MA is updated as
\begin{equation}
x_i^{(i)} \leftarrow \mv{x}^\star[i],
\end{equation}
with the positions of the remaining $(N-1)$ MAs preserved, i.e.,
\begin{equation}
x_n^{(i)} = x_n^{(i-1)}, \quad n \in \mathcal{N} \setminus \{i\}.
\end{equation}
While in the $i$-th ($i = \lfloor N/2 \rfloor + 1, \ldots, N$) iteration, the position of the $(N - i + \lfloor N/2 \rfloor + 1)$-th MA is updated as
\begin{equation}
x_{N-i+\lfloor N/2 \rfloor+1}^{(i)} \leftarrow \mv{x}^\star[N - i + \lfloor N/2 \rfloor + 1],
\end{equation}
with the positions of the remaining $(N-1)$ MAs preserved, i.e.,
\begin{equation}
x_n^{(i)} = x_n^{(i-1)}, \quad n \in \mathcal{N} \setminus \{N - i + \lfloor N/2 \rfloor + 1\}.
\end{equation}

By mathematical induction, we can show that $\mv{x}^{(i)}$ satisfies constraints \eqref{P1-1b-star} and \eqref{P1-1c-star}. For $i=0$, it is obvious that the constraints \eqref{P1-1b-star} and \eqref{P1-1c-star} hold for \(\mv{x}^{(0)} = \mv{x}\). Next, we assume that they also hold for \( \mv{x}^{(i)}, 1 \leq i \leq \lfloor N/2 \rfloor - 1 \). In the \( (i+1) \)-th iteration, we have
\begin{equation}
x_{i+1}^{(i+1)} \leftarrow \mv{x}^\star[i + 1], \;
x_n^{(i+1)} = x_n^{(i)}, \; n \in \mathcal{N} \setminus \{i + 1\}.
\end{equation}
This ensures that
\begin{align}
x_{i+1}^{(i+1)} - x_i^{(i+1)} &= \mv{x}^\star[i + 1] - x_i^{(i)} \notag \\& =\mv{x}^\star[i + 1] - \mv{x}^\star[i] = d.
\end{align}
Additionally, since
\begin{align}
x_{i+1}^{(i)} - x_i^{(i)} \geq d, \quad x_{i+2}^{(i)} - x_{i+1}^{(i)} \geq d,
\end{align}
we have
\begin{align}
x_{i+2}^{(i+1)} - x_{i+1}^{(i+1)} &=\Big(x_{i+2}^{(i+1)} - x_{i}^{(i+1)}\Big)-\Big(x_{i+1}^{(i+1)} - x_{i}^{(i+1)}\Big) \notag \\ &=\Big(x_{i+2}^{(i+1)} - x_{i}^{(i)}\Big)-d \notag \\ &=\Big(x_{i+2}^{(i)} - x_{i+1}^{(i)}\Big)+\Big(x_{i+1}^{(i)} - x_{i}^{(i)}\Big)-d \notag \\ &\geq d+d-d =d.
\end{align}
Moreover, since the positions of the remaining $(N-1)$ MAs are unchanged in the $(i+1)$-th step, i.e., $x_{n}^{(i+1)} = x_{n}^{(i)}$, $n \in \mathcal{N} \setminus \{i+1\}$, we have $x_{n}^{(i+1)} - x_{n-1}^{(i+1)} \geq d, n \in {\cal N} \setminus \{1\}$. Given that the position update process is symmetric for $i = 1,2,\ldots,\lfloor N/2 \rfloor$ and $i = \lfloor N/2 \rfloor + 1, \lfloor N/2 \rfloor + 2, \ldots, N$, it follows that constraints \eqref{P1-1b-star} and \eqref{P1-1c-star} are also satisfied for $i = \lfloor N/2 \rfloor + 1, \lfloor N/2 \rfloor + 2, \ldots, N$ via a similar procedure. Therefore, $\mv{x}^{(i)}$ satisfies constraints \eqref{P1-1b-star} and \eqref{P1-1c-star} under the proposed antenna position adjustment scheme.

Subsequently, we show that $\text{var}(\tilde{\mv{x}}^{(i)}) \geq \text{var}(\tilde{\mv{x}}^{(i-1)}), i \in {\cal N}$. Define
\begin{equation}
\tilde{\mu}(\tilde{\mv{x}},j) \triangleq \frac{1}{N-1} \sum_{n=1, n \neq j}^{N} x_n^2.
\end{equation}
Then, for $i = 1, 2, \ldots, \lfloor N/2 \rfloor$, the difference between $\text{var}(\tilde{\mv{x}}^{(i)})$ and $\text{var}(\tilde{\mv{x}}^{(i-1)})$ can be expressed as
\begin{align}
&\text{var}(\tilde{\mv{x}}^{(i)}) - \text{var}(\tilde{\mv{x}}^{(i-1)}) \notag \\
&\overset{(E_1)}{=} \frac{1}{N} \Big( \big( x_i^{(i)} \big)^4 - \big( x_i^{(i-1)} \big)^4 \Big) - \frac{1}{N^2} \bigg( \Big( (N-1)\tilde{\mu}(\tilde{\mv{x}}^{(i)}, i) \notag \\ &+\big(x_i^{(i)}\big)^2 \Big)^2 - \Big( (N-1)\tilde{\mu}(\tilde{\mv{x}}^{(i-1)}, i) + \big(x_i^{(i-1)}\big)^2 \Big)^2 \bigg) \notag \\
&\overset{(E_2)}{=} \frac{1}{N} \Big( \big( x_i^{(i)} \big)^4 - \big( x_i^{(i-1)} \big)^4 \Big) - \frac{1}{N^2} \bigg(2(N-1)\tilde{\mu}(\tilde{\mv{x}}^{(i)}, i) \notag \\ &\Big(\big(x_i^{(i)}\big)^2-\big(x_i^{(i-1)}\big)^2\Big) + \big( x_i^{(i)} \big)^4 - \big( x_i^{(i-1)} \big)^4 \bigg) \notag \\
&= \frac{N-1}{N^2} \underbrace{\left( \big( x_i^{(i)} \big)^2 - \big( x_i^{(i-1)} \big)^2 \right)}_{(a)} \notag \\ 
&\quad\underbrace{\left( \big( x_i^{(i)} \big)^2 + \big( x_i^{(i-1)} \big)^2 - 2\tilde{\mu}(\tilde{\mv{x}}^{(i)}, i) \right)}_{(b)},\label{var_diff}
\end{align}
where equality $(E_1)$ holds since $x_{n}^{(i)} = x_{n}^{(i-1)}$ for $n \in \mathcal{N} \setminus \{i\}$ and ${\mu}(\tilde{\mv{x}}^{(i)}) = \frac{1}{N} \left( (N-1)\tilde{\mu}(\tilde{\mv{x}}^{(i)}, i) + (x_i^{(i)})^2 \right)$, and equality $(E_2)$ holds because $\tilde{\mu}(\tilde{\mv{x}}^{(i-1)}, i) = \tilde{\mu}(\tilde{\mv{x}}^{(i)}, i)$.

Next, we prove that expression $(a)$ in \eqref{var_diff} is non-positive for $i = 1, 2, \ldots, \lfloor N/2 \rfloor$. Since $x_i^{(i)} = \mv{x}^\star[i] = (i-1)D$ for $i = 1, 2, \ldots, \lfloor N/2 \rfloor$, we have
\begin{align}
x_i^{(i-1)} &= \sum_{n=2}^i \big(x_n^{(i-1)}-x_{n-1}^{(i-1)}\big)+x_1^{(i-1)} \notag \\ &\overset{(I_1)}{\geq} (i-1)d = x_i^{(i)}, \quad i = 2, 3, \ldots, \lfloor N/2 \rfloor,
\end{align}
where inequality $(I_1)$ holds since $\mv{x}^{(i-1)}$ satisfies constraints \eqref{P1-1b-star} and \eqref{P1-1c-star}. For $i=1$, $x_1^{(1)}=0$ and $x_1^{(0)} \geq 0$. Hence, equation $(a)$ \eqref{var_diff} is non-positive for $i = 1, 2, \ldots, \lfloor N/2 \rfloor$.

Then, we further show that expression $(b)$ \eqref{var_diff} is also non-positive for $i = 1, 2, \ldots, \lfloor N/2 \rfloor$. Multiplying expression $(b)$ by $(N-1)$ yields
{\small
\begin{align}
& (N-1)\left(\big( x_i^{(i)} \big)^2 + \big( x_i^{(i-1)} \big)^2 - 2\tilde{\mu}(\tilde{\mv{x}}^{(i)}, i)\right) \notag \\
&= (N-1) \left(\big( x_i^{(i)} \big)^2 + \big( x_i^{(i-1)} \big)^2 \right) - 2\sum_{n=1, n \neq j}^{N} \big(x_n^{(i)}\big)^2 \notag \\ 
&= 2\sum_{n=1}^{i-1} \Big(\big(x_i^{(i)}\big)^2-\big(x_n^{(i)}\big)^2\Big) + 2\sum_{n=i+1}^{N} \Big(\big(x_i^{(i-1)}\big)^2-\big(x_n^{(i)}\big)^2\Big) \notag+ \\ & \big((N-1)-2(i-1)\big)\big(x_i^{(i)}\big)^2 +\big((N-1)-2(N-i)\big)\big(x_i^{(i-1)}\big)^2 \notag \\ 
&\overset{(I_2)}{\leq} 2\sum_{n=1}^{i-1} (i-n)d (x_i^{(i)}+x_n^{(i)}) + 2\sum_{n=i+1}^{N} (i-n)d (x_i^{(i-1)}+x_n^{(i)}) \notag \\
&=\Big((i-1)i-(N-i+1)(N-i)\Big)x_i^{(i-1)} + \Big(2\sum_{n=1}^{i-1} (i-n)d \cdot \notag \\ &x_n^{(i)} + 2\sum_{n=i+1}^{2i-1} (i-n)d \cdot x_n^{(i)}\Big) + 2\sum_{n=2i}^{N} (i-n)d \cdot x_n^{(i)} \notag \\ 
&\overset{(I_3)}{<} 2 \sum_{j=1}^{i-1} jd \cdot \Big(x_{i-j}^{(i)}-x_{i+j}^{(i)}\Big) + 2\sum_{n=2i}^{N} (i-n)d \cdot x_n^{(i)} \overset{(I_4)}{<} 0,
\end{align}}
where inequality $(I_2)$ holds for $i = 1, 2, \ldots, \lfloor N/2 \rfloor$ due to the following four steps. First,
\begin{align}
x_i^{(i)}-x_n^{(i)} &= \mv{x}^\star[i]-x_n^{(n)} = \mv{x}^\star[i]-\mv{x}^\star[n] \notag \\ 
&= (i-n)d, \quad n=1,2,\ldots,i-1.
\end{align}
Second, $x_i^{(i-1)}-x_n^{(i)} \leq (i-n)d, \;n=i+1,i+2,\ldots,N$ since
\begin{align}
x_n^{(i)}-x_i^{(i-1)} &= x_n^{(i-1)}-x_i^{(i-1)} \notag \\ &\geq (n-i)d, \quad n=i+1,i+2,\ldots,N,
\label{I_2_second}
\end{align}
for $\mv{x}$ satisfying constraints \eqref{P1-1b-star} and \eqref{P1-1c-star}. Third, $x_i^{(i)} \leq x_i^{(i-1)}, i = 1, 2, \ldots, \lfloor N/2 \rfloor$. Fourth, we have $\big((N-1)-2(i-1)\big) + \big((N-1)-2(N-i)\big)=0$. Additionally, inequality $(I_3)$ holds due to $(i-1)i-(N-i+1)(N-i) < 0$, as both $i-1<N-i+1$ and $i<N-i$ hold for $i \leq \lfloor N/2 \rfloor$. Inequality $(I_4)$ holds since the terms involving $x_n^{(i)}$ scale with $A$ and dominate the expression for a sufficiently large array aperture $A$.

Based on the above, both expressions $(a)$ and $(b)$ are non-positive for $i = 1, 2, \ldots, \lfloor N/2 \rfloor$, which leads to $\text{var}(\tilde{\mv{x}}^{(i)}) - \text{var}(\tilde{\mv{x}}^{(i-1)}) \geq 0, \;i = 1, 2, \ldots, \lfloor N/2 \rfloor$. For $i = \lfloor N/2 \rfloor + 1, \lfloor N/2 \rfloor + 2, \ldots, N$, the MA is moved towards $A$. In this stage, term $(a)$ in \eqref{var_diff} is non-negative since $x_i^{(i)} \ge x_i^{(i-1)}$. Furthermore, for a sufficiently large $A$, $x_i^{(i)}$ is close to $A$, making $(x_i^{(i)})^2$ significantly larger than the mean squared value $\tilde{\mu}(\tilde{\mv{x}}^{(i-1)}, i)$. Thus, term $(b)$ in \eqref{var_diff} is strictly positive, which leads to $\text{var}(\tilde{\mv{x}}^{(i)}) - \text{var}(\tilde{\mv{x}}^{(i-1)}) \geq 0$ in a similar manner. Therefore, for arbitrary $\mv{x}$ satisfying constraints \eqref{P1-1b-star} and \eqref{P1-1c-star}, we have $\text{var}(\tilde{\mv{x}}) = \text{var}(\tilde{\mv{x}}^{(0)}) \le \text{var}(\tilde{\mv{x}}^{(1)}) \le \cdots \le \text{var}(\tilde{\mv{x}}^{(N)}) = \text{var}(\tilde{\mv{x}}^\star)$. Hence, $\mv{x}^\star$ is an optimal solution to problem (P1-2). This completes the proof.\vspace{-9pt}

\section{Proof of Corollary \ref{Cor2}} \label{appen_Cor_2}
It follows from Theorem \ref{Th2} that the optimal APV is given by $\mv{x}^\star = [0, d, 2d, \ldots, (\lfloor N/2 \rfloor - 1)d, A - (N - \lfloor N/2 \rfloor - 1)d, A - (N - \lfloor N/2 \rfloor - 2)d, \ldots, A - d, A]^\top$. By denoting $N_l \triangleq \lfloor N/2 \rfloor$ and $N_r \triangleq N - \lfloor N/2 \rfloor$, $\text{var}(\tilde{\mv{x}}^\star)$ can be expressed as a polynomial function denoted as $f(A,N,d)$ (with the detailed expression omitted for brevity). As such, the associated worst-case CRB on the distance estimation is given by
\begin{equation}
\text{CRB}_r(\mv{x}^\star, r_{\text{max}}) = \Big(\frac{2r^2_{\text{max}}}{1-{u^\star}^2}\Big)^2 \cdot \frac{\kappa}{f(A,N,d)}.
\label{CRB_overline_r}
\end{equation}
By evaluating the sums, it can be shown that for fixed $N$ and $d$, $f(A,N,d)$ is a quartic function w.r.t. $A$, where the term $\frac{N_lN_r}{N^2}A^4$ dominates. Since $N_l, N_r \ge 1$ are independent of $A$, $f(A,N,d)$ scales as $\mathcal{O}(A^4)$ for a sufficiently large $A$. As $\text{CRB}_r(\mv{x}^\star, r_{\text{max}})$ is inversely proportional to $f(A,N,d)$, it decreases with $A$ in the order of $\mathcal{O}(A^{-4})$.\vspace{-9pt}

\section{Derivations of the CRB matrix in Case 1.3} \label{appen_1_3}
In the joint estimation of the AoA and distance for the 1D MA array via the 2D-MUSIC algorithm, the CRB matrix of the estimator vector $\mv{\eta}=[u,r]^\top$ is given by
\begin{multline}
    \text{CRB}_{\mv{\eta}}(\mv{x}, \mv{\eta}) = \frac{\sigma^2}{2}\Bigg(\sum_{t=1}^{T}\Re\bigg\{s_t^* \mv{\Psi}(\mv{x}, \mv{\eta})^\mathsf{H} \Big(\mv{I}_N-\mv{h}(\mv{x}, \mv{\eta}) \\
    \big(\mv{h}(\mv{x}, \mv{\eta})^\mathsf{H} \mv{h}(\mv{x}, \mv{\eta})\big)^{-1}\mv{h}(\mv{x}, \mv{\eta})^\mathsf{H}\Big)\mv{\Psi}(\mv{x}, \mv{\eta})s_t\bigg\}\Bigg)^{-1},
\label{1_3_app_CRB}
\end{multline}
where $\mv{\Psi}(\mv{x}, \mv{\eta})$ denotes the partial derivative matrix of the near-field steering vector $\mv{\alpha}(\mv{x}, \mv{\eta})$ w.r.t. the estimator vector, i.e.,
\begin{equation}
\small
    \mv{\Psi}(\mv{x}, \mv{\eta}) = \Big[\frac{\partial \mv{h}(\mv{x}, \mv{\eta})}{\partial u}, \frac{\partial \mv{h}(\mv{x}, \mv{\eta})}{\partial r}\Big] = [\mv{\psi}_u(\mv{x}, u),\mv{\psi}_r(\mv{x}, r)]\in \mathbb{C}^{N \times 2}.
\label{1_3_psi}
\end{equation}
By re-denoting $\text{CRB}_{\mv{\eta}}(\mv{x}, \mv{\eta})$, $\mv{h}(\mv{x}, \mv{\eta})$ and $\mv{\Psi}(\mv{x}, \mv{\eta})$ as $\text{CRB}_{\mv{\eta}}$, $\mv{\Psi}$ and $\mv{h}$, respectively, \eqref{app_a_FI} can be further expressed as
\begin{align}
    \text{CRB}_{\mv{\eta}} &= \frac{\sigma^2}{2}\Bigg(\sum_{t=1}^{T}\Re\bigg\{s_t^* \mv{\Psi}^\mathsf{H} \Big(\mv{I}_N-\mv{h} \big(\mv{h}^\mathsf{H} \mv{h}\big)^{-1}\mv{h}^\mathsf{H}\Big)\mv{\Psi}s_t\bigg\}\Bigg)^{-1} \notag \\
    &= \frac{\sigma^2}{2}\Bigg(TP \, \Re\bigg\{\mv{\Psi}^\mathsf{H}\mv{\Psi}- \frac{1}{N\lvert\beta\rvert^2}\mv{\Psi}^\mathsf{H}\mv{h}(\mv{\Psi}^\mathsf{H}\mv{h})^\mathsf{H}\bigg\}\Bigg)^{-1} \notag \\
    &= \frac{\sigma^2}{2}\Bigg(TP \, \Re\bigg\{\begin{bmatrix}
    \mv{\psi}_u^\mathsf{H}\mv{\psi}_u & \mv{\psi}_u^\mathsf{H}\mv{\psi}_r \\
    \mv{\psi}_r^\mathsf{H}\mv{\psi}_u & \mv{\psi}_r^\mathsf{H}\mv{\psi}_r
    \end{bmatrix} - \frac{1}{N\lvert\beta\rvert^2} \notag \\
    & \qquad \begin{bmatrix}
    (\mv{\psi}_u^\mathsf{H}h)(\mv{\psi}_u^\mathsf{H}h)^* & (\mv{\psi}_u^\mathsf{H}h)(\mv{\psi}_r^\mathsf{H}h)^* \\
    (\mv{\psi}_r^\mathsf{H}h)(\mv{\psi}_u^\mathsf{H}h)^* & (\mv{\psi}_r^\mathsf{H}h)(\mv{\psi}_r^\mathsf{H}h)^*
    \end{bmatrix}\bigg\}\Bigg)^{-1} \notag \\
    &= \frac{\sigma^2}{2}\Bigg(\frac{4\pi^2}{\lambda^2TPN\lvert\beta\rvert^2} \,\begin{bmatrix}
    \text{var}(\mv{\zeta}_u) & \text{cov}(\mv{\zeta}_u,\mv{\zeta}_r) \\
    \text{cov}(\mv{\zeta}_u,\mv{\zeta}_r) & \text{var}(\mv{\zeta}_r)
    \end{bmatrix}\Bigg)^{-1} \notag \\
    &= \frac{\kappa}{\text{var}(\mv{\zeta}_u)\text{var}(\mv{\zeta}_r)-\text{cov}^2(\mv{\zeta}_u,\mv{\zeta}_r)} \notag \\
    & \qquad \begin{bmatrix}
    \text{var}(\mv{\zeta}_r) & -\text{cov}(\mv{\zeta}_u,\mv{\zeta}_r) \\
    -\text{cov}(\mv{\zeta}_u,\mv{\zeta}_r) & \text{var}(\mv{\zeta}_u)
    \end{bmatrix}.
\label{app_FI_extend}
\end{align}
Therefore, the CRBs on the AoA estimation in \eqref{1_3_CRBu} and the distance estimation in \eqref{1_3_CRBr} are respectively given by
\begin{align}
    \text{CRB}_u(\mv{x}) &=\text{CRB}_{\mv{\eta}}(\mv{x}, \mv{\eta})(1,1)\notag \\
    &=\kappa \cdot \frac{\text{var}(\mv{\tilde{x}})}{\text{var}(\mv{x})\text{var}(\mv{\tilde{x}})-{\text{cov}^2(\mv{x},\mv{\tilde{x}})}},\label{app_CRBu}\\
    \text{CRB}_r(\mv{x}, \mv{\eta}) &=\text{CRB}_{\mv{\eta}}(\mv{x}, \mv{\eta})(2,2)\notag \\
    &=\kappa \cdot \frac{4r^4\text{var}(\mv{x})+8ur^3{\text{cov}(\mv{x},\mv{\tilde{x}})}+4u^2r^2\text{var}(\mv{\tilde{x}})}{(1-u^2)^2\Big(\text{var}(\mv{x})\text{var}(\mv{\tilde{x}})-\text{cov}^2(\mv{x},\mv{\tilde{x}})\Big)}.
\label{app_CRBr}
\end{align}
This thus completes the derivations.
\vspace{-9pt}

\section{Derivations of the CRB matrix in Case 2.1} \label{appen_2_1}
The CRB matrix of the estimator vector $\mv{\eta}=[u, v]^\top$ in the estimation of the target's two AoAs via the 2D-MUSIC algorithm is given by
\begin{multline}
    \text{CRB}_{\mv{\eta}}(\tilde{\mv{s}}, \mv{\eta}) = \frac{\sigma^2}{2}\Bigg(\sum_{t=1}^{T}\Re\bigg\{s_t^* \mv{\Psi}(\tilde{\mv{s}}, \mv{\eta})^\mathsf{H} \Big(\mv{I}_N-\mv{h}(\tilde{\mv{s}}, \mv{\eta}) \\
    \big(\mv{h}(\tilde{\mv{s}}, \mv{\eta})^\mathsf{H} \mv{h}(\tilde{\mv{s}}, \mv{\eta})\big)^{-1}\mv{h}(\tilde{\mv{s}}, \mv{\eta})^\mathsf{H}\Big)\mv{\Psi}(\tilde{\mv{s}}, \mv{\eta})s_t\bigg\}\Bigg)^{-1}\Bigg|_{r=r^\star},
\label{2_1_app_FIM}
\end{multline}
where $\mv{\Psi}(\tilde{\mv{s}}, \mv{\eta})$ denotes the partial derivative matrix of the near-field channel vector $\mv{h}(\tilde{\mv{s}}, \mv{\eta})$ w.r.t. the estimator vector, i.e.,
\begin{equation}
    \mv{\Psi}(\tilde{\mv{s}}, \mv{\eta}) = \Big[\frac{\partial \mv{h}(\tilde{\mv{s}}, \mv{\eta})}{\partial u}, \frac{\partial \mv{h}(\tilde{\mv{s}}, \mv{\eta})}{\partial v}\Big] \in \mathbb{C}^{N \times 2},
\label{2_3_psi}
\end{equation}
where
\begin{equation}
    \frac{\partial \mv{h}(\tilde{\mv{s}}, \mv{\eta})}{\partial u} = j\frac{2\pi}{\lambda}\mv{\xi} \odot \mv{h}(\tilde{\mv{s}}, \mv{\eta}),
\label{2_1_psi_u}
\end{equation}
and
\begin{equation}
    \frac{\partial \mv{h}(\tilde{\mv{s}}, \mv{\eta})}{\partial v} = j\frac{2\pi}{\lambda}\mv{\pi} \odot \mv{h}(\tilde{\mv{s}}, \mv{\eta}).
\label{2_1_psi_v}
\end{equation}
By following similar procedures as those in Appendix \ref{appen_1_3}, the CRBs on the two AoAs for the 2D MA array are given by \eqref{2_1_CRBu} and \eqref{2_1_CRBv}, respectively. This thus completes the derivations.
\vspace{-9pt}

\section{Derivations of the CRB in Case 2.2} \label{appen_2_2}
In estimating the target distance for the 2D MA array via the MUSIC algorithm, the Fisher information of the estimator $r$ is
\begin{multline}
    \text{CRB}_r(\tilde{\mv{s}}, r) = \frac{\sigma^2}{2}\Big(\sum_{t=1}^{T}\Re\big\{s_t^* \mv{\psi}_r(\tilde{\mv{s}}, r)^\mathsf{H} \Big(\mv{I}_N-\mv{h}(\tilde{\mv{s}}, r) \\
    \big(\mv{h}(\tilde{\mv{s}}, r)^\mathsf{H} \mv{h}(\tilde{\mv{s}}, r)\big)^{-1}\mv{h}(\tilde{\mv{s}}, r)^\mathsf{H}\Big)\mv{\psi}_r(\tilde{\mv{s}}, r)s_t\big\}\Big)^{-1}\Bigg|_{u=u^\star,v=v^\star},
\label{2_2_app_FI}
\end{multline}
where $\mv{\psi}_r(\tilde{\mv{s}}, r)$ denotes the partial derivative vector of the near-field channel vector $\mv{h}(\tilde{\mv{s}}, r)$ w.r.t. the estimator, i.e.,
\begin{equation}
    \mv{\psi}_r(\tilde{\mv{s}}, r) = \frac{\partial \mv{h}(\tilde{\mv{s}}, r)}{\partial r} = j\frac{2\pi}{\lambda}\mv{\rho} \odot \mv{h}(\tilde{\mv{s}}, r).
\label{2_2_psi}
\end{equation}
Following similar procedures as those in Appendix \ref{appen_1_1}, the CRB on the distance estimation for the 2D MA array is given by \eqref{2_2_CRBr}. This thus completes the derivations.
\vspace{-9pt}

\section{Derivations of the CRB matrix in Case 2.3} \label{appen_2_3}
In the joint estimation of the target's two AoAs and distance for the 2D MA array via the 3D-MUSIC algorithm, the CRB matrix of the estimator vector $\mv{\eta}=[u,v,r]^\top$ is given by 
\begin{multline}
    \widetilde{\text{CRB}}_{\mv{\eta}}(\tilde{\mv{s}}, \mv{\eta}) = \frac{\sigma^2}{2}\Bigg(\sum_{t=1}^{T}\Re\bigg\{s_t^* \mv{\Psi}'(\tilde{\mv{s}}, \mv{\eta})^\mathsf{H} \Big(\mv{I}_N-\mv{h}(\tilde{\mv{s}}, \mv{\eta}) \\
    \big(\mv{h}(\tilde{\mv{s}}, \mv{\eta})^\mathsf{H} \mv{h}(\tilde{\mv{s}}, \mv{\eta})\big)^{-1}\mv{h}(\tilde{\mv{s}}, \mv{\eta})^\mathsf{H}\Big)\mv{\Psi}'(\tilde{\mv{s}}, \mv{\eta})s_t\bigg\}\Bigg)^{-1},
\label{2_3_app_FIM}
\end{multline}
where $\mv{\Psi}'(\tilde{\mv{s}}, \mv{\eta})$ denotes the partial derivative matrix of the near-field channel vector $\mv{h}(\tilde{\mv{s}}, \mv{\eta})$ w.r.t. the estimator vector, i.e.,
\begin{equation}
    \mv{\Psi}'(\tilde{\mv{s}}, \mv{\eta}) = \Big[\frac{\partial \mv{h}(\tilde{\mv{s}}, \mv{\eta})}{\partial u}, \frac{\partial \mv{h}(\tilde{\mv{s}}, \mv{\eta})}{\partial v}, \frac{\partial \mv{h}(\tilde{\mv{s}}, \mv{\eta})}{\partial r}\Big] \in \mathbb{C}^{N \times 3},
\label{2_3_psi_3D}
\end{equation}
where
\begin{equation}
    \frac{\partial \mv{h}(\tilde{\mv{s}}, \mv{\eta})}{\partial r} = j\frac{2\pi}{\lambda}\mv{\rho} \odot \mv{h}(\tilde{\mv{s}}, \mv{\eta}).
\label{2_3_psi_uvr}
\end{equation}
By performing similar procedures to those in Appendix \ref{appen_1_3}, the CRBs for the two AoAs and the distance are obtained as \eqref{2_3_CRBu}, \eqref{2_3_CRBv}, and \eqref{2_3_CRBr}, respectively. This thus completes the derivations.\vspace{-6pt}

\bibliography{Ref.bib}
\bibliographystyle{IEEEtran}

\end{document}